\documentclass[fleqn,usenatbib]{mnras}

\usepackage{newtxtext,newtxmath}

\usepackage[T1]{fontenc}
\usepackage{soul}
\usepackage{libertine}

\DeclareRobustCommand{\VAN}[3]{#2}
\let\VANthebibliography\thebibliography
\def\thebibliography{\DeclareRobustCommand{\VAN}[3]{##3}\VANthebibliography}

\usepackage{graphicx}	
\usepackage{amsmath}	

\usepackage{xcolor}
\usepackage{amsfonts}
\usepackage{longtable}
\usepackage{hyperref}
\usepackage{url}
\usepackage{multirow}

\usepackage{booktabs}

\title[Constraining dark matter in neutron stars]{Constraining self-interacting fermionic dark matter in admixed neutron stars using multimessenger astronomy}

\author[M. Mariani et al.]
{Mauro Mariani,$^{1,2}$\thanks{E-mail: mmariani@fcaglp.unlp.edu.ar}
Conrado Albertus,$^{3}$
M. del Rosario Alessandroni,$^{1}$
\newauthor 
Milva G. Orsaria,$^{1,2,4}$
M. \'Angeles P\'erez-Garc\'ia,$^{3}$
and Ignacio F. Ranea-Sandoval$^{1,2}$
\\
$^{1}$Grupo de Astrof\'isica de Remanentes Compactos, Facultad de Ciencias Astron\'omicas y Geof\'isicas, Universidad Nacional de La Plata, Paseo del Bosque S/N,\\ La Plata (1900), Argentina\\
$^{2}$CONICET, Godoy Cruz 2290, 1425, CABA, Argentina\\
$^{3}$Department of Fundamental Physics and IUFFyM, University of Salamanca, Plaza de la Merced S/N E-37008, Salamanca, Spain\\
$^{4}$Department of Physics, San Diego State University, 5500 Campanile Drive, San Diego, CA 92182, USA
}

\date{Accepted XXX. Received YYY; in original form ZZZ}

\pubyear{2023}

\begin{document}
\label{firstpage}
\pagerange{\pageref{firstpage}--\pageref{lastpage}}
\maketitle

\begin{abstract}
We investigate the structure of admixed neutron stars with a regular hadronic component and a fraction of fermionic self-interacting dark matter. Using two limiting equations of state for the dense baryonic interior, constructed from piecewise generalised polytropes, and an asymmetric self-interacting fermionic dark component, we analyse different scenarios of admixed neutron stars depending on the mass of dark fermions $m_\chi$, interaction mediators $m_\phi$, and self-interacting strengths $g$. We find that the contribution of dark matter to the masses and radii of neutron stars leads to tension with mass estimates of the pulsar J0453+1559, the least massive neutron star, and with the constraints coming from the GW170817 event. We discuss the possibilities of constraining dark matter model parameters $g$ and $y \equiv m_\chi/m_\phi$, using current existing knowledge on neutron star estimations of mass, radius, and tidal deformability, along with the accepted cosmological dark matter freeze-out values and self-interaction cross-section to mass ratio, $\sigma_\mathrm{SI}/m_\chi$, fitted to explain Bullet, Abell, and dwarf galaxy cluster dynamics. By assuming the most restrictive upper limit, $\sigma_\mathrm{SI}/m_\chi < 0.1$ cm$^2$/g, along with dark matter freeze-out range values, the allowed $g$-$y$ region is $0.01 \lesssim g \lesssim 0.1$, with $0.5 \lesssim y \lesssim 200$. For the first time, the combination of updated complementary restrictions is used to set constraints on self-interacting dark matter.
\end{abstract}




\section{Introduction}
\label{sec:intro}

In addition to ordinary matter, our Universe is believed to be populated by a dark component known as dark matter (DM). Within the standard $\Lambda$-CDM cosmological model, DM accounts for approximately $80\%$ of the total matter content in the Universe. Over the past few decades, extensive searches have been conducted to find this elusive massive particle candidate with null results \citep{Bertone_2018}. The analysis of multi-messenger events involving radiation \citep{DeRocco_2019}, neutrinos \citep{rembPhysRevD.98.103010}, and gravitational waves (GW) \citep{Badurina_2021} has been explored to uncover hints of a new particle sector beyond the Standard Model.

Numerous candidates have been proposed to explain the existence of DM, including weakly (strongly) interacting massive particles, WIMPs (SIMPs), and feebly interacting neutrinos \citep{Datta:2021iot}. In recent years, significant attention has been given to exploring additional DM candidates, particularly those in the light or ultra-light mass sector, such as weakly interacting axions or axion-like particles, ALPs. Despite extensive searches, the absence of any detection has led to highly constrained parameter spaces \citep{Schumann_2019}. Axions, in particular, have gained prominence due to their potential as a solution to the charge-parity problem in Quantum Chromodynamics (QCD) \citep{Peccei:2008tsc}. They have also been investigated for their potential impact on phenomena like cooling and the generation of broadband radio signals in pulsars \citep{Prabhu_2021}. Additionally, self-interacting DM (SIDM) has been proposed as a solution to the core-cusp problem, which refers to the discrepancy between the inferred density profiles of low-mass galaxies and the predictions of cosmological N-body simulations \citep{PhysRevLett.84.3760}. SIDM has been explored as a means to reconcile these inconsistencies. These examples represent only a fraction of the wide array of DM candidates that have emerged from extensions of the Standard Model.

The clumpy nature of DM in the Universe allows for the existence of regions such as globular clusters with enhanced DM mass densities, $\rho_\chi\in [10, 10^{5}]\rho_{\chi_0}$, being $\rho_{\chi_0}\sim 0.4$ GeV/cm$^3$ the solar neighbourhood DM density \citep{Read:2014tld,Cautun:2020tmw}. Nevertheless, detecting DM particles directly is an extremely challenging task due to the fact that they rarely interact with ordinary matter. So far direct, indirect and collider searches have tried to set up different experimental strategies, in addition to gravitational effects, to discern a signal that reveals the fundamental nature of DM. However, an alternative approach lies in indirectly detecting DM through accumulation effects in astrophysical bodies and, in particular, in dense neutron stars (NSs). In this sense, astrophysical bodies characterised by mass, $M$, and radius, $R$, play a significant role in this endeavour through the compactness, defined as $C \equiv G M/(c^2 R)$, with  NSs or quark stars being especially efficient at this task \citep{MirrorSANDIN2009278, AdmixedPhysRevD.93.083009}.

Despite the extremely weak interactions between DM particles and ordinary matter, the high density of NSs enables them to effectively capture and retain DM particles that may pass through \citep{Cermeno:2015efs}. Once trapped within the star, these particles undergo successive collisions with the nucleons in the star, gradually thermalizing and eventually accumulating in the core over a sufficiently long period of time \citep{Singh:2023cpo}. This accumulation would lead to several detectable signatures, such as the heating up of cold NSs \citep{Bertone:2008csa,Bell:2018hun,ngelesPrezGarca2022CoolingON}, pulsar scintillation \citep{AngelesPerez-Garcia:2013fpo}, or changes in the rotational patterns \citep{Kouvaris:2014rja}, making them detectable in future observations. Combining the efforts in the electromagnetic bands, with infrared capabilities of James Webb Space Telescope, the proposed gamma-ray telescopes like e-ASTROGAM \citep{De_Angelis_2018} and AMEGO, the upcoming Square Kilometer Array in radio astronomy, ATHENA in X-ray astronomy, along with 3rd generation gravitational wave detectors such as advanced LIGO, Virgo, and KAGRA, Cosmic Explorer \citep{CEreitze2019cosmic} or Einstein Telescope \citep{Branchesi_2023}, represents a promising way to detect potential signals from a plethora of DM induced phenomenology \citep{Boddy:2022stf}. In extreme cases, the accumulation of DM could trigger the collapse of the star into a black hole \citep{Mcdermott:2012con, Zurek:2014adm, Singh:2023cpo} or the conversion into a quark star \citep{Herrero:2019esf}, leading to bursting radiation signals \citep{zenati2023neutrino}.

Besides accumulation in the NS core due to the stellar gravitational potential well and interactions with ordinary matter, we foresee another scenario where DM particles can appear as final or intermediate products in decay or creation processes, such as massive sterile neutrinos or massive scalars \citep{ALBERTUS2015209,rembPhysRevD.98.103010,cerdeñoPhysRevD.104.063013,fornalPhysRevLett.120.191801}. In either case, the evidence (or lack thereof) of DM accumulation in NSs can provide valuable clues on where to direct experimental efforts to detect this type of matter.

Several studies have investigated SIDM in both the weakly and the strongly interacting regimes within compact objects \citep{Narain:2006csm,leung:2011dan,Deliyergiyev:2019dco,das:2019cne,das:2021dman,Husain_2021, kain:2021dma,Das:2022dma,Miao:2022dma,Rutherford:2023cba,Routaray:2023idm,Diedrichs:2023tdo}. In this work, we model NSs including a fermionic SIDM component using a generalised piecewise polytropic (GPP) equation of state (EoS) for hadronic matter adjusted to reproduce, in a reasonably accurate way, observables like mass, radius and moment of inertia that would be obtained with realistic hadronic EoSs \citep{OBoyle:2020peo}. SIDM is incorporated through the two-fluid formalism \citep{MirrorSANDIN2009278}. 

The simplest model of SIDM particles includes a Yukawa-type potential and a force carrier $\phi$ mediating between two DM particles $\chi$, with mass $m_\chi$. The key parameter governing the likelihood of self-interactions among DM particles is the SI cross-section ($\sigma_\mathrm{SI}$) to mass ratio, $\sigma_\mathrm{SI}/m_{\chi}$. Furthermore, the separation between the DM halo of the Abell galaxy cluster and its stars can be explained in terms of self-interaction cross-sections of DM \citep[see, for example, ][{and references therein}]{Kahlhoefer:2015oti}. Additionally, SIDM provides a consistent cross-section that matches the DM halo profile of dwarf galaxies. However, there are discrepancies between this fit and the results from galaxy merger studies. This is why the analysis of velocity-dependent self-interactions related to freeze-out, $\left\langle\sigma_{\text{ann}} v_{\text{rel}}\right\rangle$ -where $\sigma_{\text{ann}}$ is the DM annihilation cross-section, and $ v_{\text{rel}}$ is the relative velocity between the annihilating particles-, which avoids the constraints posed by galaxy clusters, becomes relevant in such cases \citep{Hayashi:2021pdm}. Furthermore, it is crucial that DM self-interactions are not so strong as to disrupt the elliptical shape of spiral galaxies or displace sub-clusters, as seen, for instance, in the Bullet Cluster. Moreover, the latter can also be employed to establish restrictions on the SIDM cross-section \citep{Robertson:2016wdt}. All these constraints place limits on the strength of DM self-interactions; for more details, see \cite{cosmoSI10.1093/mnras/sts514}.

On the other hand, astronomical observations of NS received a boost in the past two decades. Detection in double pulsar systems -PSR J1614-2230 \citep{Demorest:2010bx}, PSR J0348+0432 \citep{Antoniadis:2013pzd}, and PSR J0740+6620 \citep{Cromartie:2020rsd, Fonseca:2021rfa}- of the $\sim 2\,M_\odot$ NSs requires an acceptable EoS able to support such high masses. These observations posed the first strong restrictions to the behaviour of matter inside such compact stars. Moreover, multimessenger astronomy with GWs detected from the binary NS merger event GW170817 and its electromagnetic counterpart allowed to restrict the value of the dimensionless tidal deformability, $\Lambda$, of NSs \citep{TheLIGOScientific:2017qsa,Annala:2017llu,Most:2018hfd,Raithel:2018tdf,Abbott:2018exr,Capano:2019eae} and ejecta properties \citep{P_rez_Garc_a_2022}. Additionally, the Neutron Star Interior Composition Explorer (NICER) has measured the mass and radius of the isolated millisecond-pulsars PSR J0030+0451 \citep{Riley:2019yda,Miller:2019cac} and (together with XMM-Newton data) PSR J0740+6620 \citep{Riley:2021pdl,Miller:2021qha} with great precision. The latter study showed that despite having a mass $\sim 40\, \%$ larger, the radii of PSR J0740+6620 and PSR J0030+0451 are of the same order \citep{Riley:2021pdl,Miller:2021qha}.

Our goal is to analyse how the current astrophysical constraints associated with NSs and cluster dynamics can set limits on the mass of the DM particle, $m_{\chi}$, and on parameters of the SIDM model, such as the self-interaction (SI) coupling strength constant, $g$, or the mass scale of the interaction or, equivalently, the associated generic mediator mass, $m_{\phi}$. Furthermore, for the first time, constraints coming from multi-messenger astronomy of NSs are used, combined with the restrictions of SI cross-section from galaxy clusters, dwarf galaxies and the thermally averaged annihilation cross-section related to the cosmological DM freeze-out value, to set constraints to SIDM.

The paper is organised in the following manner. Section \ref{section2} is devoted to describing the theoretical framework used in this paper. In particular, Subsection \ref{subsection2a} provides the description of both the hadronic EoS and fermionic SIDM EoS used in this study. Subsection \ref{subsec:twofluid} offers a brief review of the two-fluid formalism employed to investigate NSs with a DM component using the two aforementioned EoS. Our main findings are presented in Section \ref{sec:results}, while Section \ref{sec:conclusions} contains a summary and a discussion of the astrophysical implications of our results. Unless stated otherwise, we use units where $G=c=\hbar=1$ throughout the paper.

\section{Theoretical Framework} \label{section2}

\subsection{Generic hadronic matter and fermionic SIDM EoS}
\label{subsection2a}

First, let us discuss the description of the hadronic content in NSs. To ensure that our conclusions are not dependent on any specific EoS, we constructed two different hadronic EoSs using the GPP formalism presented in \cite{OBoyle:2020peo}.

These two EoSs, referred to as the \textit{soft} and \textit{stiff}, are designed so that the resulting mass-radius relationships encompass a family of EoSs consistent with chiral effective field theory (EFT) up to a mass density of $\rho=1.1 \,\rho_{\rm nuc}$, being $\rho_{\rm nuc}=0.16$~fm$^{-3}$ the nuclear saturation density \citep{Hebeler:2013nza, Annala:2020efq}. In both cases, the crust is described by a GPP fit to the SLy4 crust EoS, presented in \cite{OBoyle:2020peo}. Within these parameterised EoSs, the pressure $p(\rho)$, energy density $\epsilon(\rho)$, and speed of sound $c_s(\rho)$ are continuous functions. This is a crucial aspect of the GPP formalism as the dimensionless tidal deformability explicitly depends on the value of the speed of sound \citep[for more details, see, for example, ][and references therein]{leung:2022tdo}.

For each interval in mass density range $[\rho_{i-1},\rho_i]$ the EoS adopts a power-law form
\begin{eqnarray}
p(\rho) &=& K_i \rho^{\Gamma_i}+\Lambda_i \, , \\
\epsilon(\rho) &=& \frac{K_i}{\Gamma_i-1} \rho^{\Gamma_i}+(1+a_i) \rho-\Lambda_i \, ,   
\end{eqnarray}
where the parameters $K_i, \Gamma_i, a_i, \Lambda_i$ characterise the fit and ensure continuity and differentiability of both $p(\rho)$ and $\epsilon(\rho)$ at the dividing densities $\rho_0$,  $\rho_1$ and $\rho_2$, where $\rho_0 < \rho_1 < \rho_2$. This leads to continuous speed of sound, $c_{\rm s}$. By imposing subsequent mathematical relations among these quantities, we obtain seven parameters used to construct each hadronic EoS, as presented in Table~\ref{tabla:param_eos_had}.

The \textit{soft} and the \textit{stiff} EoSs act as an envelope for numerous microscopic hadronic EoSs in the literature that satisfy current astrophysical constraints on NSs. This approach has recently been used to study both macroscopic and oscillating properties of compact objects independently of any particular theoretical model \citep[see, for example, ][]{Ranea:2022bou,goncalves:2022ios,saes:2022eos, lugones:2023ama,Lenzi:2023hsw,Ranea:2023auq,ranea-sandoval:2023cmr}.

\begin{table}
\addtolength{\tabcolsep}{-2pt}
\centering
\begin{tabular}{cccccccc}
\toprule
 & $\log_{10}\rho_0$ & $\log_{10}\rho_1$ & $\log_{10}\rho_2$  & $\Gamma_1$ & $\Gamma_2$ & $\Gamma_3$ & $\log_{10}K_1$ \vspace{0.1cm}  \\
\midrule
\textit{soft} & 13.990 & 14.31 & 14.40 & 2.750 & 6.20 & 2.9 & -27.33 \\
\textit{stiff} & 13.902 & 14.05 & 14.78 & 2.764 & 3.17 & 2.5 & -27.22 \\
\bottomrule
\end{tabular}
\caption{Parameters of the selected hadronic EoSs constructed with the prescription described in \protect\cite{OBoyle:2020peo}.}
\label{tabla:param_eos_had}
\end{table}

In addition to considering hadronic matter, we also incorporate an additional fraction of massive fermionic DM particles that can inhabit the NS. This arises in our scenario as a result of the dark component being gravitationally bound to the compact star.

The capability of a NS to capture DM from an external distribution is mainly determined by its gravitational pull, opacity, and the kinematics of incoming particles from the surrounding dark environment \citep{press1985ApJ...296..679P, gould1987ApJ...321..571G}. 
It is important to note that the DM capture rates in NSs have been calculated based on scattering off nucleons ($N$) \citep{kouvarisPhysRevD.77.023006}. Typically, the capture rate $C_\chi$ can be approximated as follows -see Eq.~($1$) in \cite{cermeño_pérez-garcía_silk_2017}-,
\begin{equation}
    C_\chi \simeq 1.8 \times 10^{25}\left(\frac{1 \,\mathrm{GeV}}{m_{\chi}}\right)\left(\frac{\rho_{\chi}}{\rho_{\chi_0}}\right) 
    \nu_\chi \; \mathrm{s}^{-1},
\end{equation}
where $\nu_\chi$ denotes the probability of at least one scattering event between a DM particle ($\chi$) and a $N$ taking place within the NS. At this point, it is worth emphasising that in our treatment the \mbox{$\chi$-$N$} interaction is considered secondary with respect to gravitational effects. The previous expression considers NS mass and radius values, along with General Relativity corrections, for typical benchmark compact stars with masses around $M\sim 1.4 M_\odot$ and radii of approximately $R\sim 10$ km. This is accurate to within factors of order unity, and any associated uncertainties are inherently included in the fraction of DM populating the star.

Setting a minimum upper limit on the DM-nucleon cross-section, \mbox{$\sigma_{\chi N}\gtrsim 10^{-46}\,\rm cm^2$,} results in $\nu_\chi \sim 1$, indicating that the NS saturates its capability to capture and bind DM. Therefore, the amount of DM inside the NS is a tiny fraction compared to the baryonic number, which is of the order of $10^{58}$. Given the current rates and the fact that the oldest NS lifetimes are $\tau\sim 10^9$ yr, it is likely that an additional mechanism is necessary to incorporate DM to the few-per-cent level, as assumed in most works, including ours.

Now, using a simplified scattering model, we will consider the contributing terms for the $\chi \chi \rightarrow \chi \chi$ process. We will assume that the SI terms are described by generic Lorentz scalar $S$ and vector $V_\mu$ mediator fields. Explicitly, the Lagrangian density for DM interaction involving $\chi$, DM anti-particles $\bar{\chi}$ and $S$, $V_\mu$  fields is written as
\begin{equation}
\mathcal{L}_{\mathrm{SIDM, int}} \supset- g_V \bar{\chi} \gamma^\mu \chi V_\mu + g_S \bar{\chi} S\chi ,
\end{equation}
where $g_S$ and $g_V$ are the scalar and vector coupling strengths, respectively, and $\gamma ^\mu$ are the Dirac matrices. Generically, we assume a bosonic mediator, $\phi$, which we can take to be either a Lorentz scalar or a vector to illustrate, with mass scale $m_\phi \sim m_I$, where $m_I$ is the interaction mass scale. In its simplest form, this scenario assumes a unified fine-structure constant $\alpha_\chi\sim \alpha_{S,V}$ (where $\alpha_{S,V}$ are, respectively, the scalar and vector fine structure constants) and $m_\phi \ll m_\chi$. Since we consider asymmetric DM coupled to light $S,\,V,$ force mediators, if they are sufficiently light, then the interaction between DM particles becomes long-range. More specifically, for an interaction described in the non-relativistic regime by a sort of Yukawa potential, $\mathcal{V}_{S,V}=-\alpha_{S,V} e^{-m_{S,V} r} / r$, where $\alpha_{S,V}=g_{S,V}^2/4\pi$, long-range effects occur if the mediator mass is smaller than the Bohr momentum, $m_{S,V} \lesssim \alpha_{S,V} m_{\chi}/2$.

Following the work of \cite{AdmixedPhysRevD.93.083009}, we model SIDM as a Fermi gas of SI particles with mass $m_\chi$. The explicit expressions for the fermionic SIDM EoS are given by \cite{AdmixedPhysRevD.93.083009},
\begin{eqnarray}
P_{\chi}&=& \frac{m_\chi^4}{24 \pi^2} \left[(2\zeta_\chi^3-3\zeta_\chi) \sqrt{1+\zeta_\chi^2} \right. \nonumber \\ 
&& \left. + 3 \ln{\left(\zeta_\chi+\sqrt{1+\zeta_\chi^2}\right)} \right]  
+ \left(\frac{m_\chi^2}{3\pi^2} \right)^{2} y^2 \zeta_\chi^6 \,, \\
\epsilon_{\chi}&=&2\,\left(\frac{y\, n_\chi}{m_\chi}\right)^{2}+n_\chi\, m_\chi\, \sqrt{1+\zeta_\chi^2}-P_{\chi} \,.
\label{eqeps}
\end{eqnarray}

In order to handle dimensionless quantities, we have defined \mbox{$\zeta_\chi=k_{F,\chi}/m_\chi$} as the ratio of DM Fermi momentum to the DM particle mass and a strength parameter $y=m_\chi/m_I$, the quotient between the DM particle mass and the interaction mass scale. The terms that are proportional to $y^2$ account for the SI effects, which are included analogously to the vector interaction terms in Nambu--Jona-Lasino-type models for quark matter \citep{Orsaria:2013hna}. In our fermionic SIDM model, we can approximate the self-scattering cross-section $\sigma_\mathrm{SI}$ using dimensional arguments \citep{Girmohanta:2022csc}. This allows us to express the self-interaction terms in Eq.~\eqref{eqeps} in the following form
\begin{equation}
    \epsilon_{\chi\chi}=2\left(\frac{y n_\chi}{m_\chi}\right)^2-\left(\frac{m_\chi^2}{3 \pi^2}\right)^2 y^2 \zeta_\chi^6=\left(\frac{y n_\chi}{m_\chi}\right)^2 \, .
\end{equation}

Since we do not expect large values of accreted or produced DM in the interior of the NS and assuming a two-body self-interaction for simplicity, we note that the contribution to the energy density or pressure is proportional to the square of the scalar density $n_{s,\chi}$ or the vector density $n_\chi$. Note that $n_\chi=\rho_{\chi}/m_\chi$ is the $\chi$ number density, and in this context,
\begin{equation*}
    n_\chi=  \frac{k_{F,\chi}^3}{3 \pi^2 } \, . 
\end{equation*}

For \mbox{$\zeta_\chi=k_{F,\chi}/m_\chi \rightarrow 0$}, we find that $n_{s,\chi}$ can be approximated by
\begin{equation*}
    n_{s,\chi}=\frac{1}{\pi^2} \int_0^{k_{F,\chi}} \frac{m_\chi k^2}{\sqrt{k^2+m_\chi^{ 2}}} dk \approx \frac{k_{F,\chi}^3}{3 \pi^2 }+O\left(\zeta_\chi^5\right) \, .
\end{equation*}

Thus, in our scenario, $n_{s,\chi}\simeq n_{\chi}$. These scalar and vector particle number densities contribute with the same strength to $\epsilon_{\chi}$ and  $P_{\chi}$, affecting the DM EoS. 

In the non-relativistic limit, the scalar SI cross-section for \mbox{$\chi \chi \rightarrow \chi \chi$} scattering can be expressed as
\begin{equation}
    \sigma_{S,\chi \chi \rightarrow \chi \chi}=\frac{m_\chi^2 g_S^4}{8 \pi m_S^4} \, .
\end{equation}

Therefore, the ratio of the SI cross-section to the mass of the $\chi$ particle becomes
\begin{equation}
   \frac{ \sigma_{S,\chi \chi \rightarrow \chi \chi}}{m_\chi}=\frac{y_S^4 g_S^4}{8 \pi m_\chi^3},
\end{equation}
where we have introduced the notation $y_S=m_\chi/m_S$. It is worth noting that in the non-relativistic limit, a vector interaction yields the same expression for the SI cross-section to mass ratio, i.e., \mbox{$\sigma_{S,\chi \chi \rightarrow \chi \chi}\sim \sigma_{V,\chi \chi \rightarrow \chi \chi}$}. Henceforth, we will consider a generic coupling constant $g_S = g_V = g$ and impose a single mass scale \mbox{$m_S = m_V = m_\phi = m_I$} to constrain the product $yg$ or equivalently $g/m_I$, based on the combined properties of NS mass-radius, dimensionless tidal deformability, $\Lambda$, thermal DM freeze-out value and the SI cross-section obtained from galactic dynamics.

In the current DM paradigm, to achieve a finite DM relic density, it is necessary to consider the self-annihilation processes involving $\bar{\chi}$ fields. This introduces a correction to the scattering SI cross-section, which can be expressed in a generic form as
\begin{equation}\label{eq:sigma_si}
 \frac{\sigma_{\mathrm{SI}}}{m_\chi}=\frac{3 y^4 g^4}{16 \pi m_\chi^3} \, .
\end{equation}

Therefore, it is important to assess the extent to which the DM in our SIDM model is consistent with the observed DM relic abundance, thereby determining the value of thermally averaged cross-section $\left\langle\sigma_{\text{ann}} v_{\text{rel}}\right\rangle$. The commonly used canonical value for a generic weakly interacting DM candidate is typically stated as \mbox{$\langle\sigma_{\mathrm{ann}} v_{\mathrm{rel}}\rangle \sim 3 \times 10^{-26}$ $\rm cm^3s^{-1}$}, with unspecified uncertainty, and is assumed to be independent of the $\chi$ mass. Recent studies on the search for annihilation products of DM suggest that \mbox{$2.2 \times 10^{-26}$~cm$^3$/s $\lesssim \langle \sigma_\mathrm{ann} v_\mathrm{rel}\rangle \lesssim 5.2 \times 10^{-26}$~cm$^3$/s}, with a weak dependence on $m_\chi >10$~GeV \citep{steigman_PhysRevD.86.023506}. The irreducible annihilation channel $\chi \chi \rightarrow \phi \phi$ and tree-level cross-section reads
\begin{equation}\label{eq:freeze_out}
    \langle\sigma_{\mathrm{ann}} v_{\mathrm{rel}}\rangle =\frac{\pi \alpha_\chi^2}{m_\chi^2} \, .    
\end{equation}

It should be noted that this expression can be modified by $\mathcal{O}(1)$ pre-factors, which we will neglect at this point, depending on factors such as whether $\chi$ is a Majorana or a Dirac fermion and whether $\phi$ is a scalar or a gauge boson. Additionally, even in simple models, $\langle \sigma_{\mathrm{ann}} v_{\text{rel}}\rangle$ typically receives contributions from other annihilation channels. In our scenario, this is described by the process $\bar{\chi} \chi \rightarrow S\,S$. At the lowest order, it can be approximated as
\begin{equation}\label{eq:freeze_out_sim}
    \left\langle\sigma_{\text{ann}} v_{\text{rel}}\right\rangle \sim \frac{3 g^4 }{128 \pi m_\chi^2} \, ,
\end{equation}
where $\sigma_{\mathrm{ann}}=\sigma_{\bar{\chi} \chi \rightarrow S\,S}$ explicitly. Furthermore, in the context of the process $S \rightarrow \bar{\chi} \chi$, the decay width is given by 
\begin{equation}\label{eq:Gamma_decay}
    \Gamma_{\phi \rightarrow \bar{\chi} \chi}=\frac{g^2}{\pi m_\phi^2}\left(\frac{1}{4} m_\phi^2-m_\chi{ }^2\right)^{\frac{3}{2}} \, .    
\end{equation}
where $\phi \equiv S$. To ensure perturbative consistency, it is necessary to verify that $\Gamma_{S \rightarrow \bar{\chi} \chi}/m_\phi \sim \delta \ll 1$. By considering $\delta \sim 0.1$, we find that $g\lesssim 1.8$, allowing for a positive and loosely constrained coupling strength. Note that for $y>1/2$ this bound is still valid as derived in \cite{Peskin:1995ev}.

\begin{figure}
\centering
\includegraphics[width=.98\linewidth]{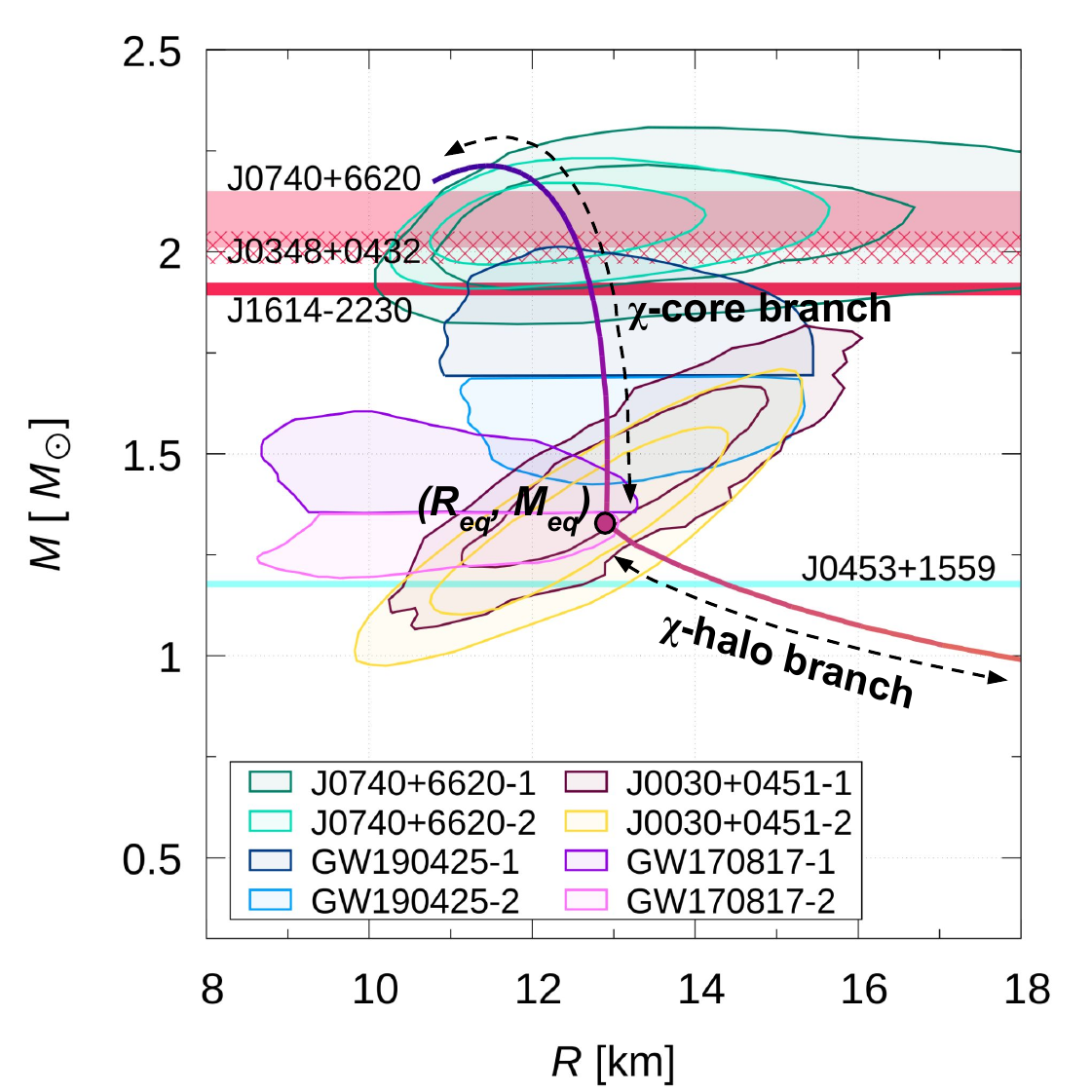}
\caption{Schematic mass-radius relationship for a typical admixed NS with a SIDM component. We indicate over the curve the two possible branches:  the $\chi$-halo branch corresponds to the configurations with a DM halo, where \mbox{$R_\chi>R_\mathrm{mat}$}, while the $\chi$-core branch includes those with a DM core, where \mbox{$R_\chi<R_\mathrm{mat}$.} The limiting configuration between branches is indicated with a circle, where $R_{\mathrm{eq}}=R_\chi=R_\mathrm{mat}$.  In the cases in which $R_{\mathrm{eq}}$ exists, \mbox{$M_{\mathrm{eq}}=M(R_{\mathrm{eq}})$}. The coloured bars and contoured clouds correspond to astrophysical constraints for NSs (see the text for details).}
 \label{fig:mraio_sch}
\end{figure}

\begin{figure*}
\centering
\includegraphics[width=0.49\linewidth]{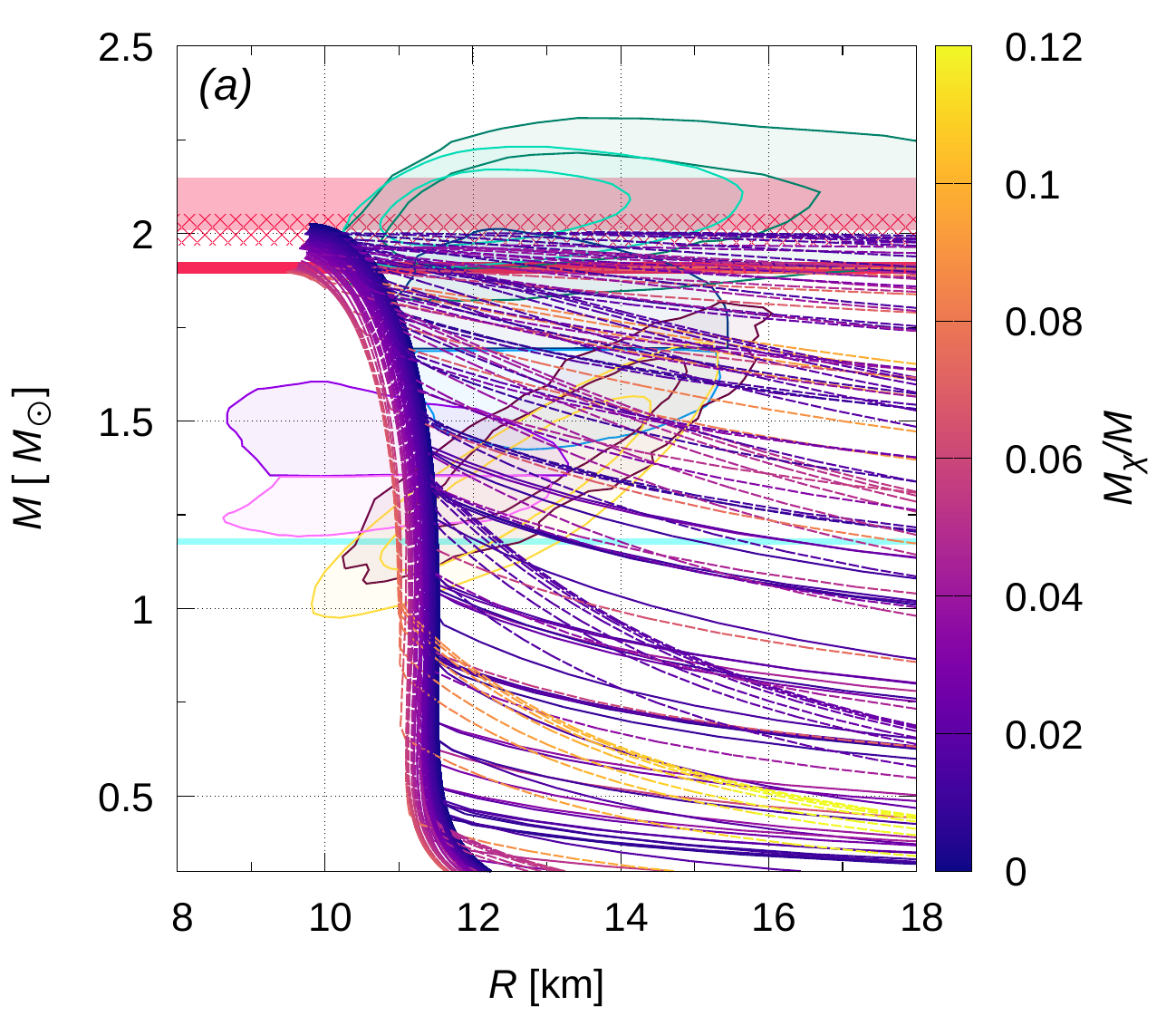}
\includegraphics[width=0.49\linewidth]{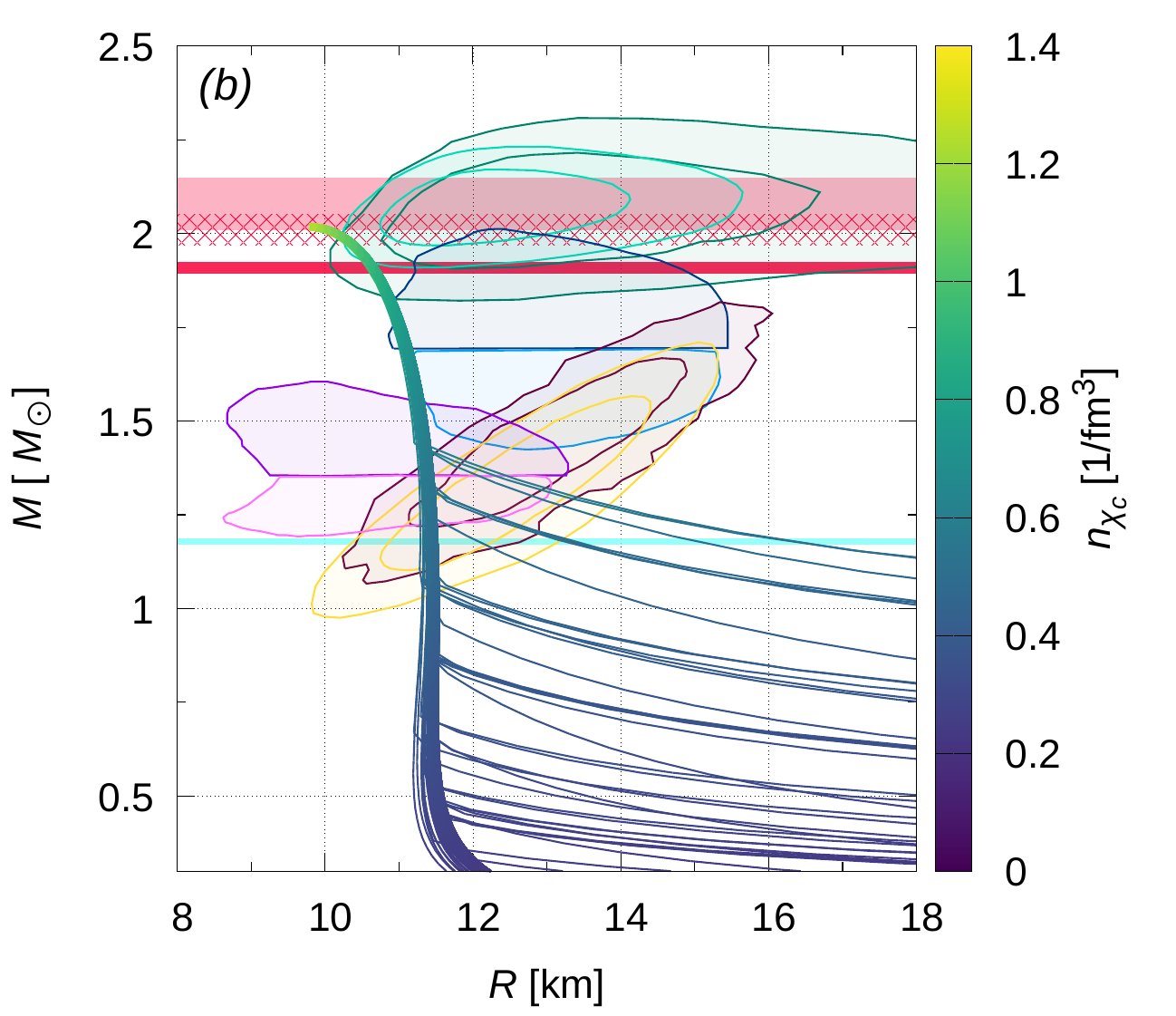}
\caption{Mass-radius relationship of different families of \textit{soft} admixed  NSs. Panel $(a)$ shows the results with all the considered EoSs. Panel $(b)$ displays the filtered EoS that fulfil the astrophysical constraints. The colour on the curves represents the value of the ratio $M_\chi/M$ (left panel) and of the central $\chi$ number density, $n_{\chi_c}$ (right panel) for each star. In both panels, it can be seen the diverse morphology of the $\chi$-halo branches and the variation of the $M_{\mathrm{eq}}$. In panel (a), it is evident that increasing $M_\chi/M$ reduces the maximum mass and radii in the $\chi$-core branches. In panel (b), similar to the effect of increasing the central baryon density, as central $n_{\chi_c}$ increases, the mass also increases. The coloured bars and contoured clouds represent the current astrophysical constraints; see Fig.~\ref{fig:mraio_sch} and text for more details. The inclusion of the SIDM in the NS models may satisfy these constraints or result in the exclusion of certain baryonic EoSs (see discussion in the text). Note that solutions in this plot, by construction, do always include a finite fraction of DM.}
 \label{fig:mraio_soft}
\end{figure*}

\subsection{Stellar configurations in the two fluid formalism}\label{subsec:twofluid}

Considering that DM interacts only weakly with ordinary matter, it is possible to obtain equilibrium configurations of compact objects formed by an admixture of ordinary and DM in the so-called two-fluid formalism in which hadronic (mat) and dark ($\chi$) components interact only gravitationally (see \citet{AdmixedPhysRevD.93.083009} and references therein). Thus, in this model, the pressure and energy density for each type of matter are assumed to be essentially decoupled, so that pressure can be written as follows
\begin{equation}
    P(r)=P_{\rm mat}(r)+P_\chi(r) \, ,
\end{equation}
and energy density can be expressed as
\begin{equation}
    \epsilon(r)=\epsilon_{\rm mat}(r)+\epsilon_\chi(r) \, .
\end{equation}

Since in this formalism, the DM-ordinary matter interaction is assumed to be negligible, and both components interact only gravitationally, the classical Tolman–Oppenheimer–Volkoff (TOV) equations are replaced by the following four coupled differential equations for pressure and gravitational mass
\begin{eqnarray}
\frac{dP_{i}(r)}{dr} &=& -\frac{M(r)\epsilon_i(r)}{r^2} \left(1 + \frac{P_{\rm i}(r)}{\epsilon _{\rm i} (r)}\right) \\ \nonumber
&& \times \left( 1 + 4\pi r^3 \frac{P(r)}{M(r)}\right) \left(1 - \frac{2M(r)}{r}\right)^{-1}, \\
\frac{dM_{i}(r)}{dr} &=& 4\pi r^2 \epsilon _{\rm i} (r) \, ,
\end{eqnarray}
where $i=\mathrm{mat},\chi$. Note that, in this description, the total gravitational mass is defined as \mbox{$M(r)= M_{\rm mat}(r) + M_{\chi}(r)$}.  The admixed TOV equations are simultaneously solved numerically using the prescribed EoSs of Table~\ref{tabla:param_eos_had} for ordinary matter and the DM EoS described in Section \ref{section2}. Along with the specified EoSs, it is necessary to make explicit the boundary conditions at the centre of the star, $r=0$, setting \mbox{$M_{\rm mat}(0) = M_{\chi}(0) = 0$.} We also need to specify the central pressures of the two fluids. For this purpose, we define the central pressure of the matter, $P_{\rm mat}(0)$, and the DM fraction, $f_\chi$, which is given by
\begin{equation}
    f_\chi= \frac{P_\chi(0)}{P_\chi(0)+P_{\rm mat}(0)}\,,
    \label{fchi}
\end{equation}
from which the central value of the dark fluid pressure, $P_\chi(0)$, can be determined.

The radius of the stellar configuration is defined as \mbox{$R=\max(R_{\rm mat}, R_\chi)$} where the radius of each component is determined when the condition $P_i(R_i)=0$ is satisfied. Due to the monotonicity of the radial coordinate, a stellar configuration could exist on the stable curve of the admixed TOV solutions in which dark and ordinary matter have the same size, i.e., $R_{\mathrm{eq}}=R_\chi=R_\mathrm{mat}$. For configurations with $R>R_{\mathrm{eq}}$ a DM halo forms, and for $R<R_{\mathrm{eq}}$ there is a DM core. In such case, if there is a stable configuration with $R=R_{\mathrm{eq}}$, its gravitational mass can be defined as $M_{\mathrm{eq}}=M(R_{\mathrm{eq}})$.

\section{Results}\label{sec:results}

To obtain the main results of this work, some of the SIDM EoS space parameters are considered within the following ranges. The DM particle mass is explored in the region
\begin{displaymath}
    100~\mathrm{MeV} \le m_\chi \le 10^4~\mathrm{MeV} \, ;
\end{displaymath}
while the mediator mass, whose value for our fermionic SIDM model is not determined, lies in the interval
\begin{displaymath}
    1~\mathrm{MeV} \le m_\phi \le 500~\mathrm{MeV} \, 
\end{displaymath}
in line with \cite{Bramante:2014bos}, corresponding to an interaction parameter $0.2 \lesssim y \lesssim 10^4$. Moreover, in accordance with the studies conducted by \cite{Panotopoulos:2017roo,Lopes:2018dma,Baryakhtar:2017dkh,Das:2022dma, Miao:2022dma}, we consider three distinct  values for the contribution of DM to the pressure at the center of the star,
\begin{displaymath}
    f_{\chi} = 0.02,\, 0.05,\, 0.1 \, .
\end{displaymath}

Within these parameter ranges, we construct a variety of admixed NSs using the previously mentioned \textit{soft} and \textit{stiff} hadronic EoSs. The results for the mass-radius plane, dimensionless tidal deformability, second Love number $k_2$, and DM parameter spaces $m_{\phi}$-$m_{\chi}$ and $g$-$y$ are presented in the following subsections.

\begin{figure*}
\centering
\includegraphics[width=0.49\linewidth]{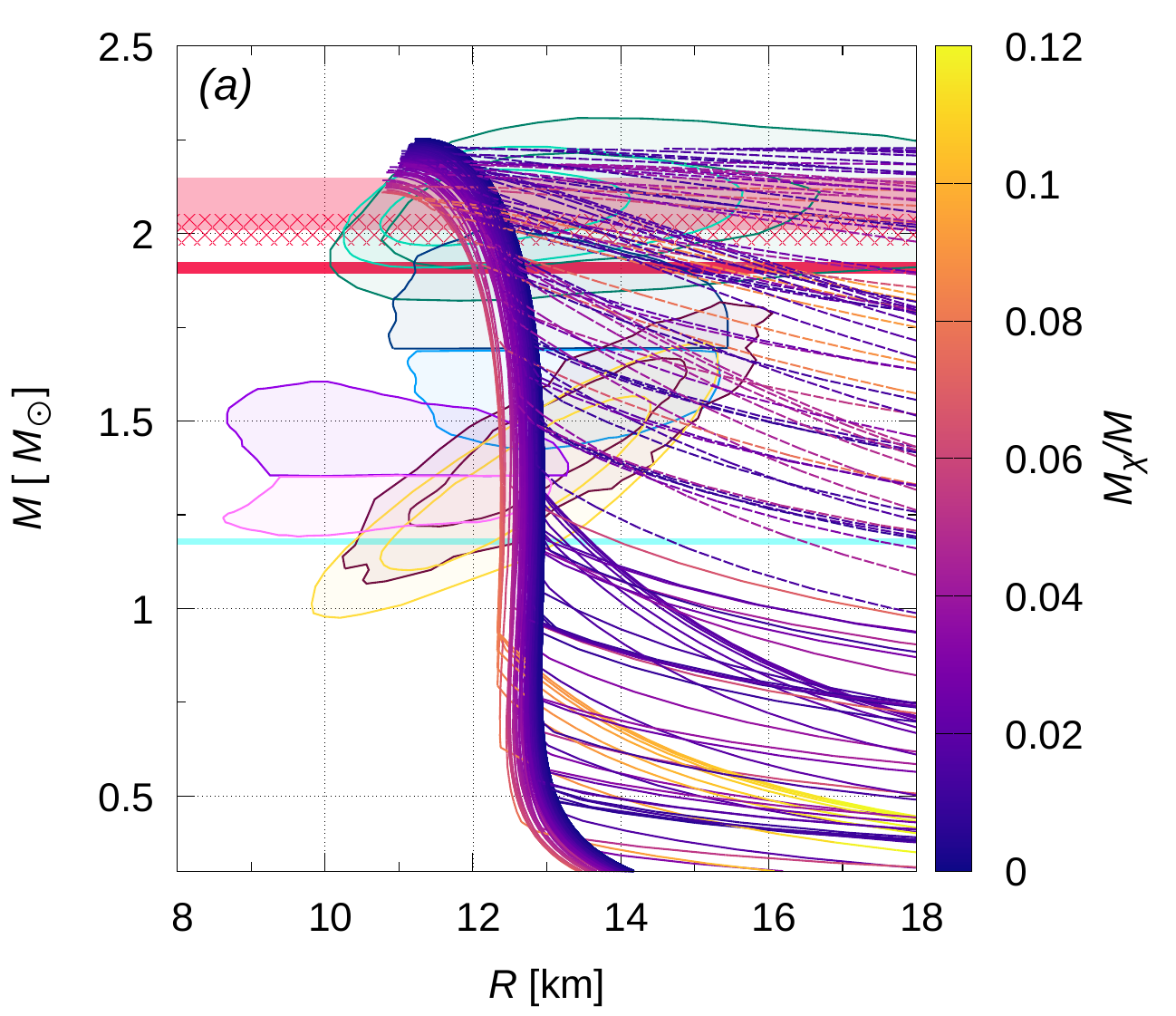}
\includegraphics[width=0.49\linewidth]{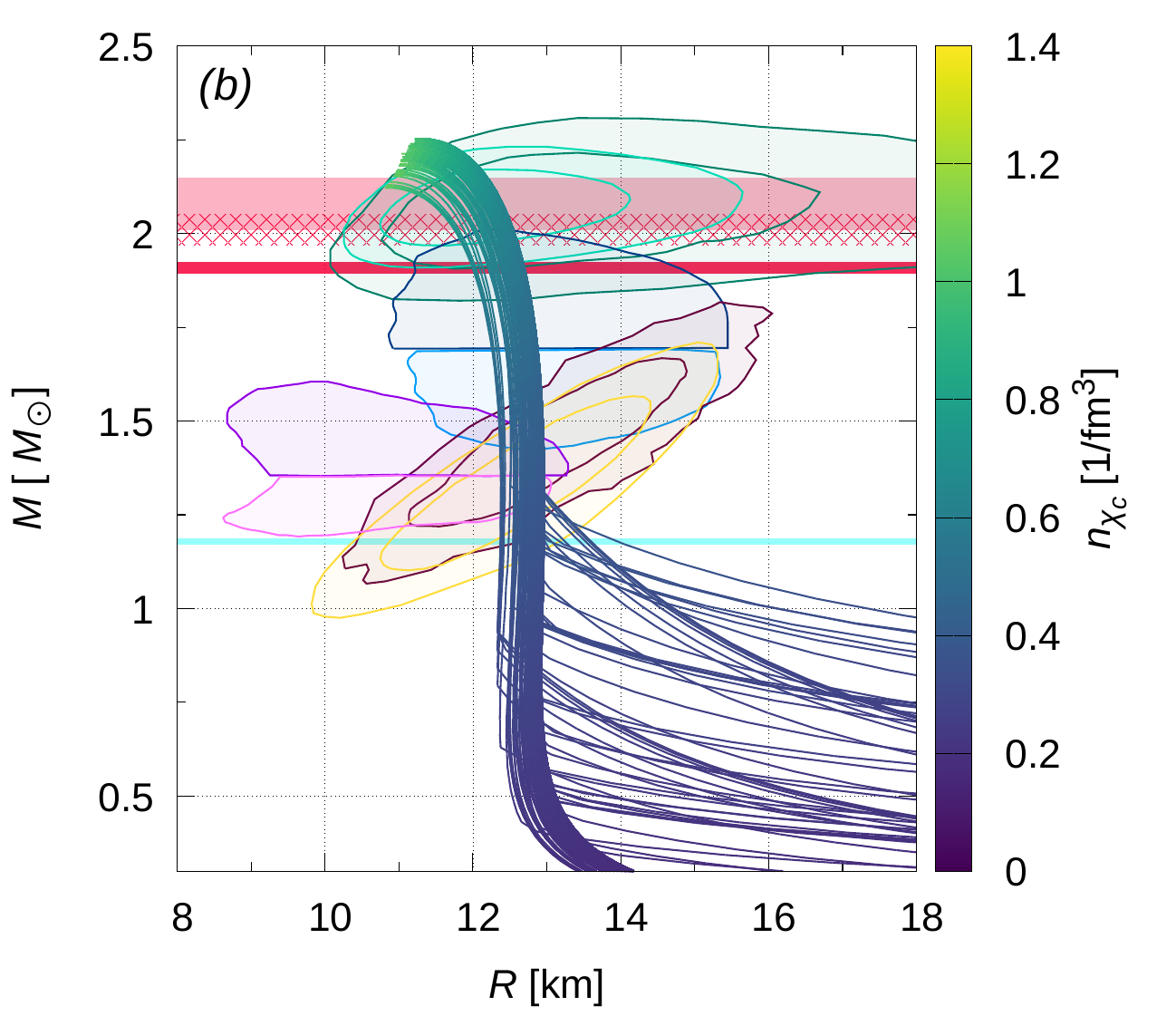}
\caption{Same as Fig.\ref{fig:mraio_soft} but for \textit{stiff} admixed  NSs.}
 \label{fig:mraio_stiff}
\end{figure*}

\subsection{Mass-radius curve for admixed NSs with fermionic SIDM}

First, in Fig.~\ref{fig:mraio_sch}, we present a schematic mass-radius relationship, illustrating an example of an admixed NS with a DM component. In the diagram, we define several relevant quantities: the $\chi$-halo branch, corresponding to configurations with a DM halo, $R_\chi>R_\mathrm{mat}$; the $\chi$-core branch, corresponding to configurations with a DM core, $R_\chi<R_\mathrm{mat}$, and the limiting stellar configuration between these branches, denoted by $R_{\mathrm{eq}}=R_\chi=R_\mathrm{mat}$ and $M_\mathrm{eq}=M(R_\mathrm{eq})$. While the $\chi$-core branch generally exhibits the typical morphology of hadronic NSs, the $\chi$-halo branch deviates from this shape, showing a sudden change of $dM/dR$ beyond $R_\mathrm{eq}$ toward larger radii. In this figure, we also present in detail the current astrophysical constraints. The bars correspond to mass constraints derived by different pulsar observations: solid and hatched red bars correspond, respectively, to pulsars J1614-2230 and J0348+0432, originally measured by \cite{Antoniadis:2013pzd} and \cite{Demorest:2010bx}; the constraint for pulsar J1614-2230 was subsequently improved in 2018 by \cite{Arzoumanian:2018tny}. The light red solid bar corresponds to the restriction in the mass of the pulsar J0740+6620 \citep{Cromartie:2020rsd}, later refined by \citet{Fonseca:2021rfa}. The cyan bar is associated with the least massive neutron star in a double pulsar system known to date, J0453+1559, with a mass of \mbox{$M = 1.174 \pm 0.004 M_\odot$} \citep{martinez:2015pja}. The coloured contoured clouds correspond to $M$ and $R$ determinations by GW or X-ray observations. The pink and purple clouds correspond to constraints derived from the observations of the GW170817 event, at $90\%$ confidence\citep{TheLIGOScientific:2017qsa, Abbott:2018exr}; the dark and light blue clouds correspond to constraints from the GW190425 event, also at $90\%$ confidence \citep{gw190425-detection}. Both GW events are displayed for the low-spin prior scenario. The brown \citep{Miller:2019cac} and yellow \citep{Riley:2019yda} clouds correspond to the restrictions derived from NICER observations of the pulsar J0030+0451. The dark green \citep{Miller:2021qha} and light green \citep{Riley:2021pdl} clouds represent the constraints obtained from the joint observation of PSR J0740+6620 by NICER and XMM-Newton. For both PSR J0030+0451 and PSR J0740+6620, two contoured clouds are presented (labelled in the figure as $1$ and $2$) since both observations were analysed and reduced by two independent research groups, yielding different results. In these last two constraints, the outer (inner) contours correspond to $95\%$ ($68\%$) confidence level. All current astrophysical constraints detailed above should be satisfied by admixed NSs.

The results for the mass-radius of the admixed NSs are shown in Fig.~\ref{fig:mraio_soft} -for the \textit{soft} hadronic EoS- and in Fig. \ref{fig:mraio_stiff} -for the \textit{stiff} hadronic EoS-. In both figures, panels $(a)$ displays all the studied EoS -considering the entire mentioned ranges of the $m_\chi$, $m_\phi$ and $f_\chi$ parameters-, while panels $(b)$ show only the results satisfying all the astrophysical constraints that we will now detail. The impact of including DM in the stellar configurations has been analysed by imposing the three most relevant astrophysical constraints as filters to obtain our results: the lower limit of the most massive double pulsar measured to date $M_\mathrm{max} \gtrsim 2.01\,M_\odot$ (light red horizontal bar) \citep{Fonseca:2021rfa}, the upper limit to the mass of PSR J0453+1559, the less massive NS in a binary system measured to date, $M_\mathrm{min} < 1.178\,M_\odot$ (cyan horizontal bar) \citep{martinez:2015pja}, and the data coming from the binary NS merger associated to the GW170817 event (purple and pink clouds) \citep{TheLIGOScientific:2017qsa,Abbott:2018exr}. Our results reveal that these three constraints are the most restrictive ones. Once these three are met, all other current astrophysical constraints for NSs are satisfied; in this sense, the other constraints are encompassed by these three filters. After applying the constraint filters, we discarded all the EoS either having too massive $\chi$-halo branch, producing \mbox{$M_\mathrm{min} > 1.178 \,M_\odot$} or laying over the GW170817 clouds, or yielded an excessive amount of DM, resulting in $M_\mathrm{max}<2.01\,M_\odot$. The colour of the curves in panel (a) represents the $M_\chi/M$ ratio for each stellar configuration, while panel (b) displays the central DM particle number density, $n_{\chi_c}$. 

\begin{figure*}
\centering
\includegraphics[height=0.35\linewidth]{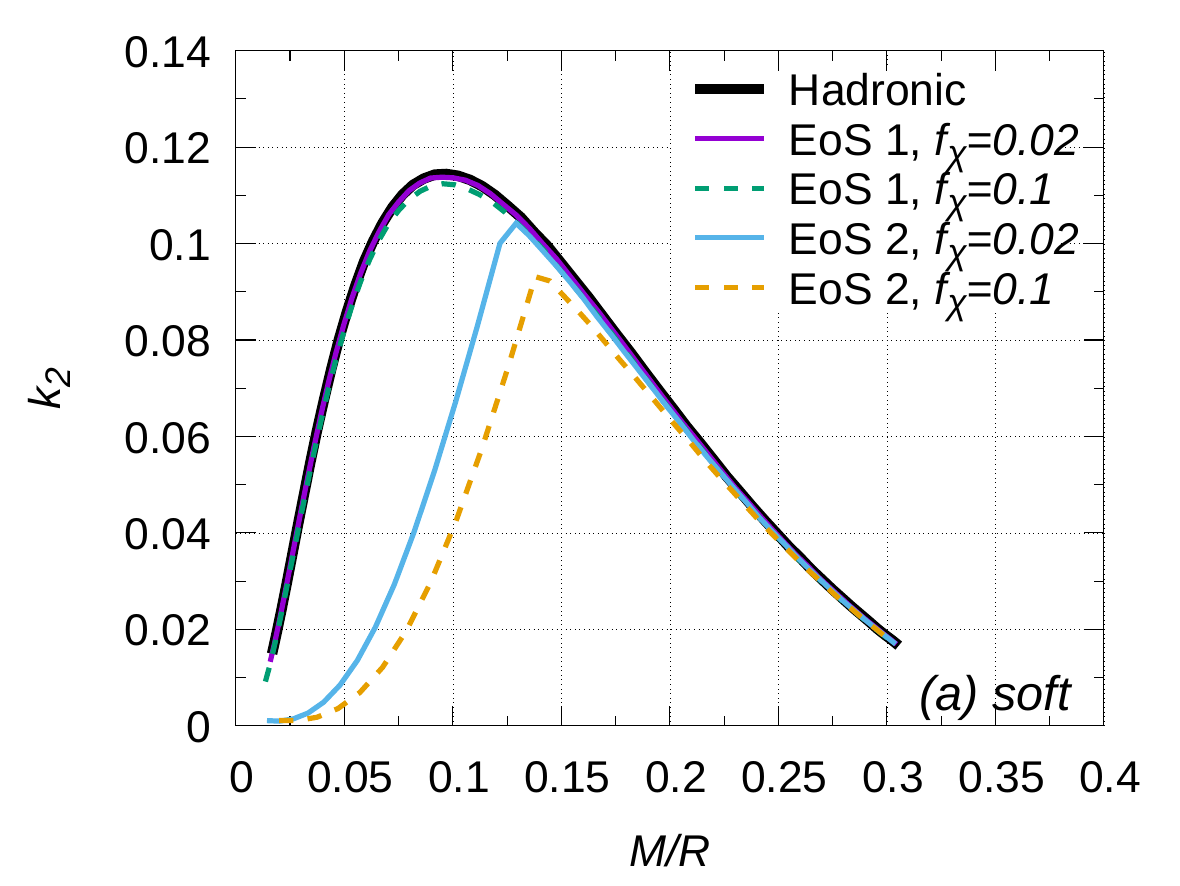}
\includegraphics[height=0.35\linewidth]{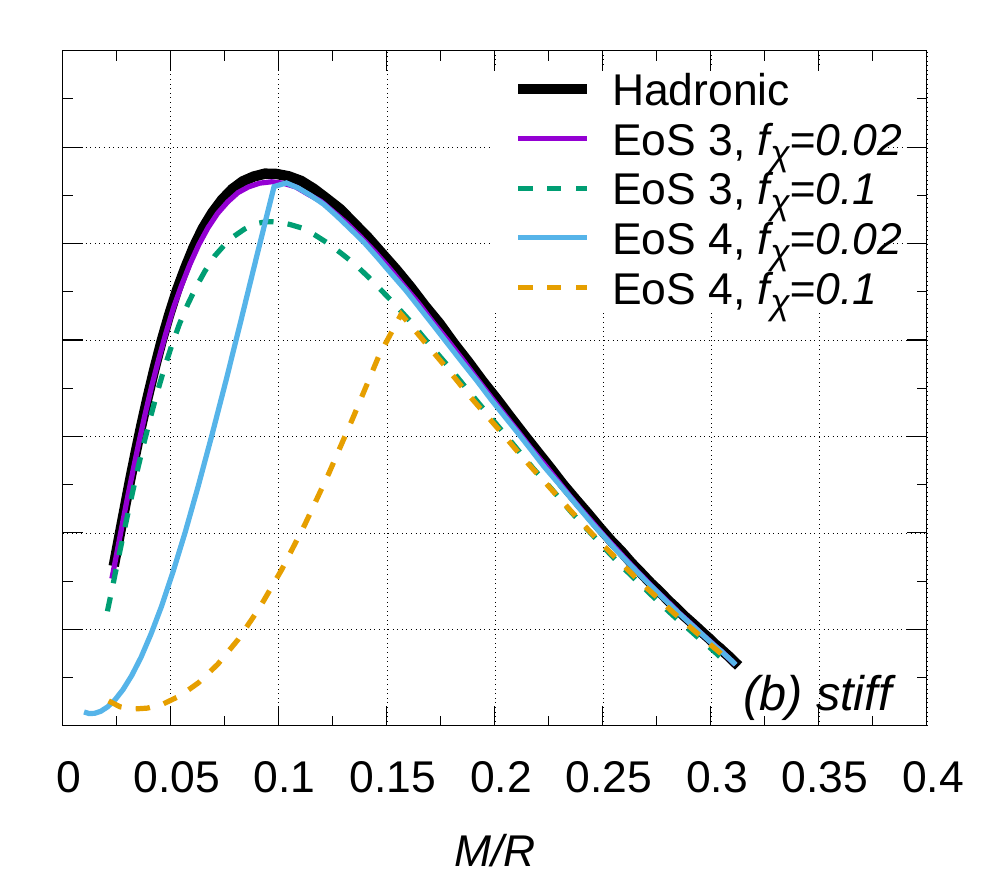}
\caption{Love number, $k_2$, as a function of the compactness for the \textit{soft}, panel $(a)$ and \textit{stiff}, panel $(b)$ hadronic EoSs of the Table~\ref{table:k2}. For completeness, the black curves correspond to purely hadronic EoSs. While the similarities or differences of the pure $\chi$-core configurations with the hadronic EoS depend on the particular selection of the parameters, the curves with significant $\chi$-halo branches present an important deviation from the hadronic curve.}
\label{fig:k2}
\end{figure*}

Although the wide array of curves displays tangled behaviour, each panel presents a variety of curves that show different features in our model results. In each panel we observe the existence of distinct $\chi$-core branches, reminiscent of typical purely hadronic mass-radius curves, but with varying arc length extensions. Thus, as the $\chi$-mass component increases, the maximum mass and radii decrease. On the other hand, the most notable effect of the DM population is the emergence of the $\chi$-halo branches that detach from the $\chi$-core branches. These  $\chi$-halo branches exhibit a more horizontal trend towards larger radii with minimal mass variation. The value of $M_\mathrm{eq}$ where these branches join also depends strongly on the model parameters and we will discuss these dependencies in Subsection~\ref{sec:parameter-space}. Nevertheless, it worth be noting here that the values of $M_\mathrm{eq}$ do not appear to correlate with the $M_\chi/M$ ratio, as there is a significant variation in the $M_\mathrm{eq}$ values unrelated to the increase or decrease of $M_\chi/M$. In particular, for the \textit{soft} model cases, an increase in the $M_\chi/M$ leads to a decrease in the maximum mass value, resulting in the exclusion of most of the curves with larger values of $M_{\chi}/M$ when the filter associated with the most restrictive mass constraint $M_\mathrm{max} \geq 2.01\,M_\odot$ is applied \citep{Fonseca:2021rfa}, panel $(b)$ of Fig.~\ref{fig:mraio_soft}. Although there are some stellar configurations for some particular EoS that contain $M_\chi/M \gtrsim 0.1$, the majority of EoSs, as well as all stellar configurations with astrophysical interest, that is, with $ M_\mathrm{min} \gtrsim 1 \,M_\odot$, have $M_\chi/M < 0.06$. As mentioned earlier, the increasing of DM in the $\chi$-core objects produces a smaller radius. This effect implies that larger fractions of DM could lead to extremely stiff hadronic EoSs that satisfy the constraint related to  GW170817 event. Considering the opposite case, \textit{soft} models results suggest that for high fractions of DM, some baryonic EoSs could be excluded since they would fail to satisfy the GW190425 constraint (dark-blue and light-blue clouds in the figures).

Regarding the behaviour of $n_{\chi_c}$ in the panel $(b)$ of both figures, it is important to emphasise that we integrate the admixed TOV equations while considering the relationship between the central values of both the hadronic and DM EoS through the parameter $f_\chi$. Consequently, it is expected that configurations with higher mass will be obtained as the central density $n_{\chi_c}$ -along with $n_{\textrm{mat}_c}$- increases, similar to what occurs in the pure hadronic scenario. In both the \textit{soft} and \textit{stiff} models, there are a few particular EoSs that suggest that the GW170817 event could potentially be produced by \mbox{$\chi$-halo} objects. 

\subsection{Effect of fermionic SIDM on NS tidal polarisability}

It is known that tidal effects in the late inspiralling phase of a binary NS merger could be detected by GW detectors. To describe this phenomenology the individual NS tidal deformability is defined as $\Lambda=\frac{2}{3} k_{2} C^{-5}$ with $k_{2}$ the second Love number \citep{Hinderer2008}. In Fig.~\ref{fig:k2}, we present the Love number $k_2$ as a function of the compactness, $C$, for \textit{soft}, panel $(a)$, and \textit{stiff}, panel (b), hadronic EoSs, using the selection of Table~\ref{table:k2}. We also include the results for purely hadronic EoS for comparison. We selected four particular admixed EoSs to consider and compare combinations of the \textit{soft} and the \textit{stiff} scenarios with two of them (EoS~$1$ and $3$) having no $\chi$-halo branch, and the other two (EoS~$2$ and $4$) featuring a substantial presence of a $\chi$-halo branch.

\begin{table}
\centering
\begin{tabular}{cccc}
\toprule
EoS & Hadronic & m$_{\chi}${[}MeV{]} & m$_{\phi}${[}MeV{]} \\
\midrule
1 & \multirow{2}{*}{soft}  & 3.0 $\times$ 10$^3$  & 2.1 $\times$ 10$^2$ \\ 
2 &                        & 2.1 $\times$ 10$^2$  & 1.5 $\times$ 10$^1$ \\ 
\midrule
3 & \multirow{2}{*}{stiff} & 1.3 $\times$ 10$^3$  & 1.5 $\times$ 10$^2$ \\ 
4 &                        & 3.0 $\times$ 10$^2$  & 3.0 $\times$ 10$^1$ \\ 
\bottomrule
\end{tabular}
\caption{EoSs chosen to analyse the behaviour of the second Love number, $k_2$, as a function of the compactness of admixed NSs. EoS~$1$ and $3$ have an absence of $\chi$-halo branch, while EoS~$2$ and $4$ have a significant presence of a $\chi$-halo branch.} 
\label{table:k2}
\end{table}

As observed in Fig.~\ref{fig:k2}, the cases with $\chi$-halo branches exhibit significant differences from the purely hadronic curve, consistently being smaller. However, the extent of the differences between the latter and the cases where the $\chi$-halo branch is absent depends on the specific admixed EoS chosen. This suggests that the study of DM through the determination of $k_2$ is influenced by the underlying hadronic models, although certain trends can still be identified. This becomes evident when comparing the results of the \textit{soft} EoS~$1$ and the purely hadronic EoS, as they are indistinguishable from each other, as is shown in panel~$(a)$. Furthermore, this characteristic appears to be independent of the DM fraction $f_\chi$. In our setting, the SIDM model is more sensitive to smaller values of DM particles in the region where $C\lesssim 0.15$.

\begin{figure*}
\centering
\includegraphics[height=0.42\linewidth]{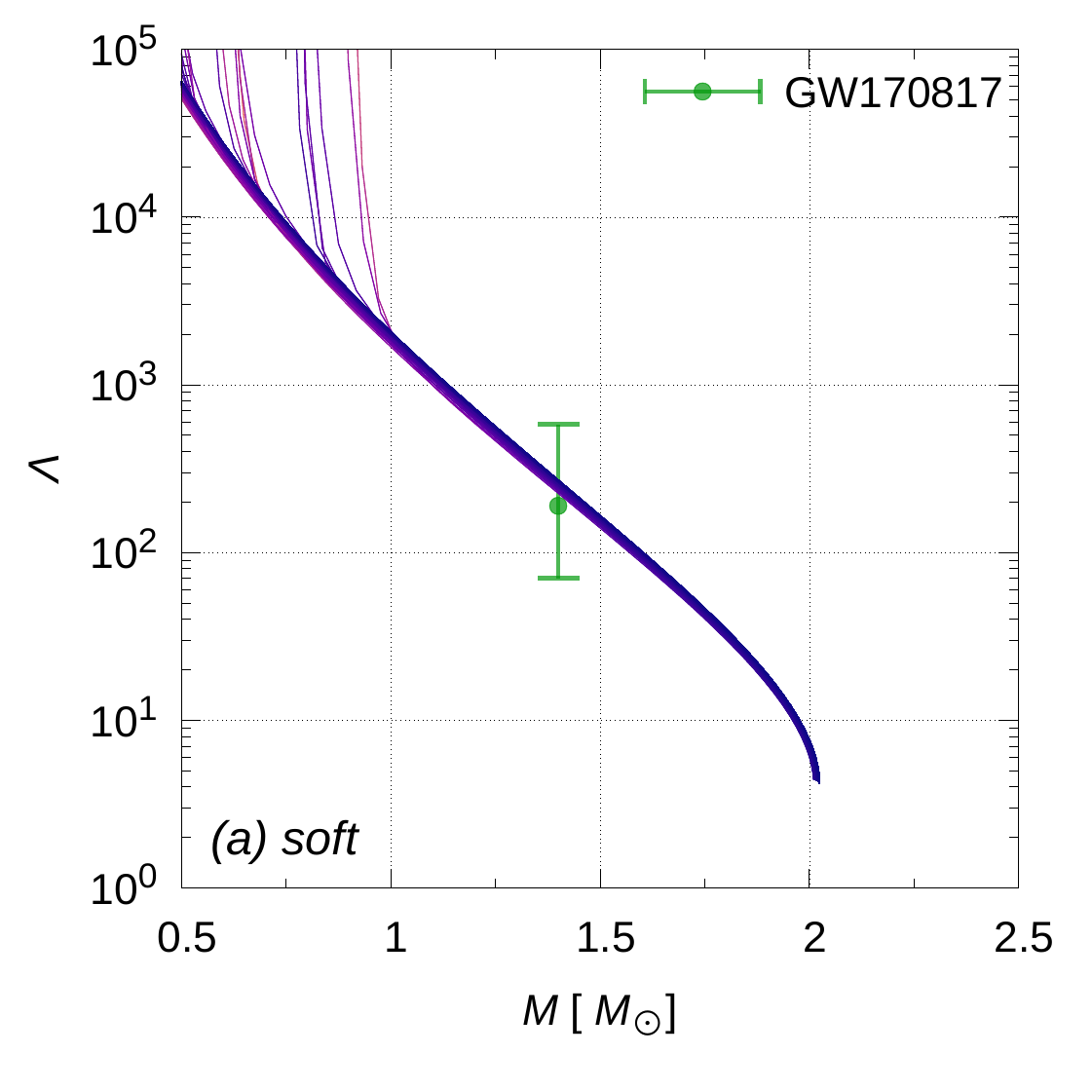}
\includegraphics[height=0.42\linewidth]{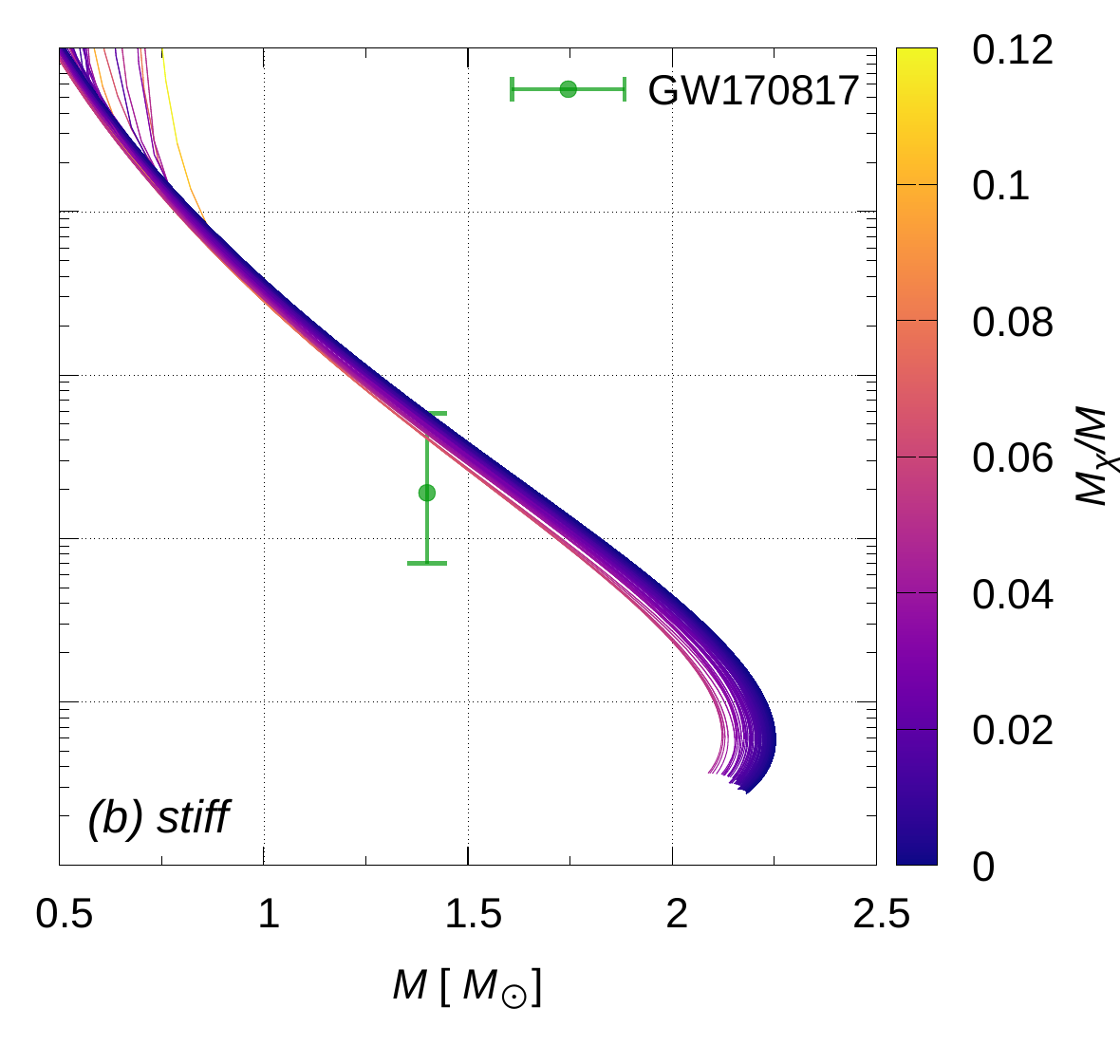}
\caption{$\Lambda$-mass plane for the \textit{soft}, panel $(a)$ and \textit{stiff}, panel (b), admixed NSs satisfying the astrophysical NSs constraints, panels $(b)$ of Figs.~\ref{fig:mraio_soft} and \ref{fig:mraio_stiff}. The colour of the curves represents the ratio between the DM mass and the total gravitational mass, $M_\chi/M$. The nearly vertical branches in the curves with a higher value of $\Lambda$ correspond to configurations where a DM halo exists, $R_\chi>R_\mathrm{mat}$. The branches displaying a traditional hadronic morphology correspond to configurations with a DM core, $R_\chi<R_\mathrm{mat}$. The green vertical segment corresponds to the constraint imposed for a $1.4 \, M_\odot$ NS based on the analysis of the GW170817 event \citep{Abbott:2018exr}. Similar to the effect of $M_\chi/M$ observed on Figs.~\ref{fig:mraio_soft} and \ref{fig:mraio_stiff}, as shown in panel $(b)$, the inclusion of DM component may contribute to satisfying the dimensionless tidal deformability constraint for some of the otherwise discarded EoSs.}
 \label{fig:tidal}
\end{figure*}

On the other hand, the constraints from GW170817 event are related to the total gravitational mass of the binary system, \mbox{$M \approx 2.74 M_{\odot}$}, and the individual NS gravitational masses $M_1 \approx (1.36-1.6) M_{\odot}$ and \mbox{$M_2 \approx(1.17-1.36) M_{\odot}$}. An analysis was performed for this event, which was marginalised over the selection methods \citep{Abbott:2018exr}, leading to the discovery of an upper bound for the effective dimensionless tidal deformability of the binary, $\tilde{\Lambda} \leq 800$ at a $90\%$ confidence level. This analysis assumed a low-spin prior, which disfavours EoSs predicting larger radii for stars. These findings arise due to the individual $\Lambda_1,\Lambda_2$ that are in-built in the actual definition of $\tilde{\Lambda}(M_1,M_2,\Lambda_1, \Lambda_2)$ \citep{Abbott:2018exr}. Tighter constraints were found in a follow-up reanalysis \citep{PhysRevX.9.011001} obtaining $\tilde{\Lambda}=300_{-230}^{+420}$ (using the $90\%$ highest posterior density interval), under minimal assumptions about the nature of the compact star binary system. In our work, we use a more refined value of the tidal deformability of a $1.4 \,{M}_{\odot}$ NS obtained from \cite{Abbott:2018exr} and estimated to be \mbox{$\Lambda_{1.4}=$ $190_{-120}^{+390}$} at a $90\%$ credible level when a common EoS is imposed. Note that this value is actually more constraining than the one obtained when assuming a binary NS merger involving similar merging NSs, in which case $\tilde{\Lambda}=\Lambda$.

The results for the $\Lambda$-mass plane are presented in Fig.~\ref{fig:tidal}, for the \textit{soft}, panel $(a)$ and \textit{stiff}, panel (b), hadronic EoS. In these figures, we only show the filtered family of stars satisfying the astrophysical constraints, corresponding to panels $(b)$ of Figs.~\ref{fig:mraio_soft} and \ref{fig:mraio_stiff}. The colour of the curves represents the ratio between the DM mass and the total gravitational mass of the respective star, denoted by $M_\chi/M$. It can be observed that \textit{soft} models satisfy the GW170817 constraint (indicated by the green segment) regardless of the DM parameters. For \textit{stiff} models, configurations with a small proportion of DM, resembling the purely hadronic configuration, barely satisfy this constraint.  Moreover, as the DM ratio increases, the GW170817 constraint is easily satisfied, consistent with the behaviour observed in the mass-radius plane, where the radius decreases with increasing DM ratio.

\begin{figure*}
\centering
\includegraphics[width=1\linewidth]{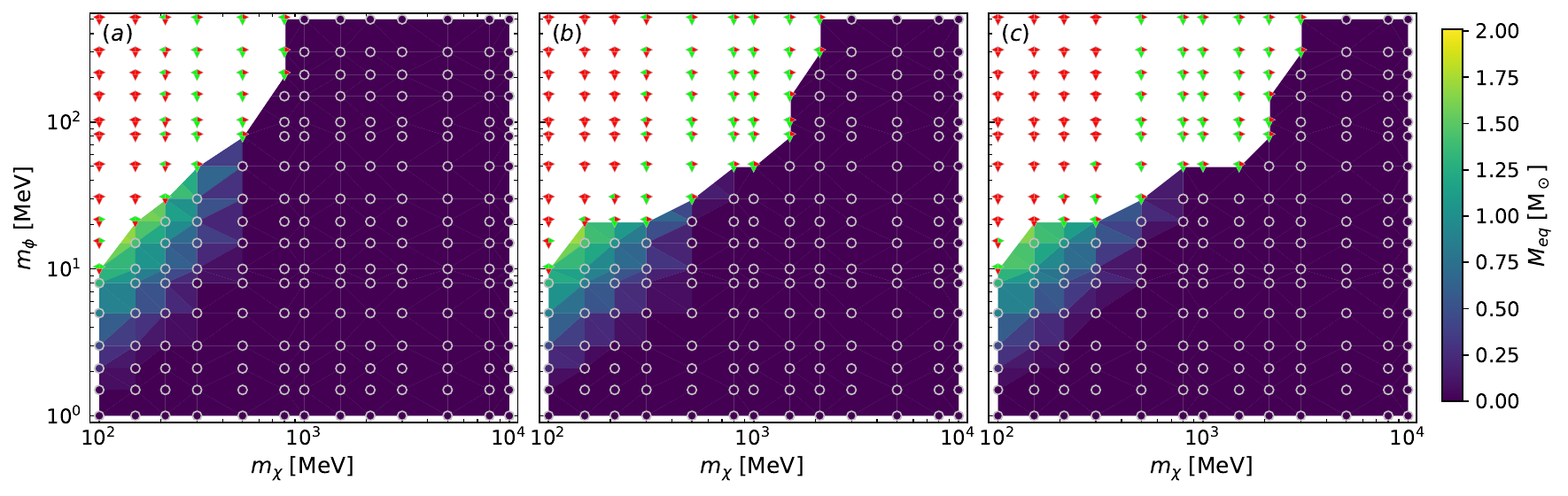}
\caption{Plane of $m_{\phi}$-$m_{\chi}$ for \textit{soft} hadronic EoSs and for different values of $f_\chi=0.02, 0.05, 0.1$ -panels $(a)$, $(b)$ and $(c)$, respectively-. The circles and diamonds indicate different EoSs within the ranges of the SIDM parameters chosen. In each panel, circles (diamonds) indicate the stellar configurations that do (do not) fulfil the three main astrophysical constraints considered (see text for details). The colour map in the circle regions indicates the value of the mass $M_{\rm eq}$ for the stellar configuration with $R_\chi=R_\mathrm{mat}$ (see details in the text). The three-sided diamonds indicate which of the three constraints is (in green) or is not (in red) satisfied. More specifically, the right and left sides of each diamond correspond to the NS maximum and minimum mass restrictions, respectively; the lower side of each diamond corresponds to the limitations imposed by GW170817.}
\label{fig:mapa_soft}
\end{figure*}

\begin{figure*}
\centering
\includegraphics[width=1\linewidth]{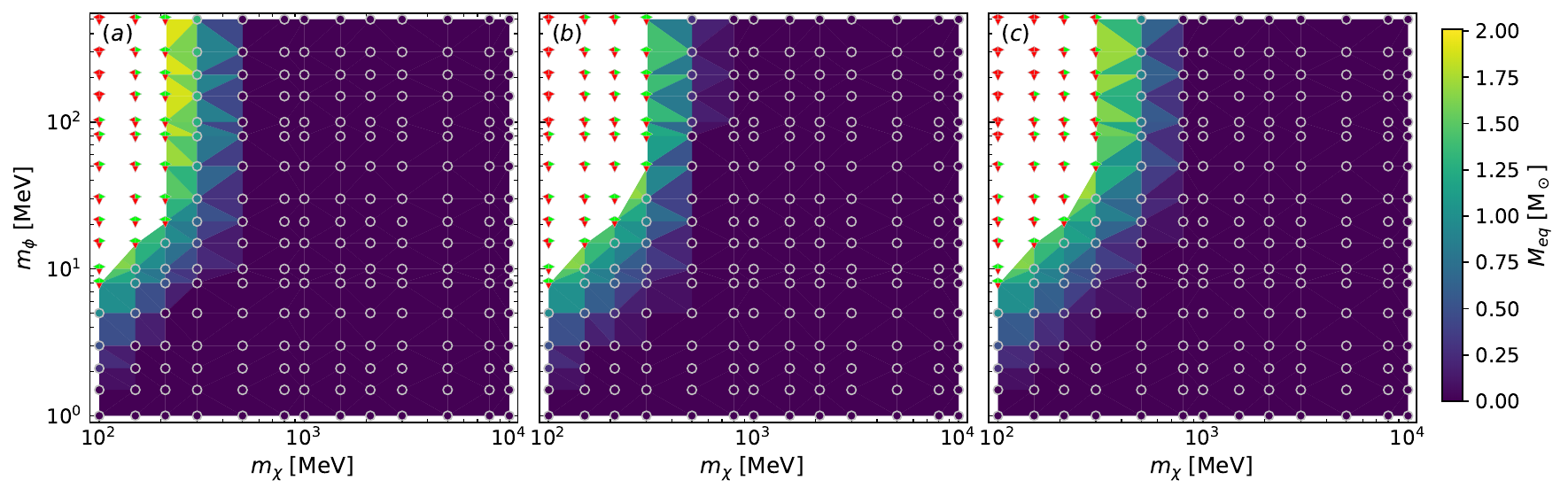}
\caption{Same as Fig.~\ref{fig:mapa_soft} but for \textit{stiff} hadronic EoSs.}
 \label{fig:mapa_stiff}
\end{figure*}

\subsection{Analysis and constraints on the SIDM-model parameter space}\label{sec:parameter-space}

After integrating the two-fluid TOV equations for the admixed EoSs, we study how applying the astrophysical filters mentioned above constrains the parameter space of SIDM. In Figs.~\ref{fig:mapa_soft} and \ref{fig:mapa_stiff} for the \textit{soft} and \textit{stiff} hadronic EoSs, respectively, we present the plane \mbox{$m_\phi$-$m_\chi$} for different values of $f_\chi=0.02, 0.05, 0.1$, within the allowed mass ranges for $m_\phi$ and $m_\chi$. The circles and diamonds indicate the sampling sets studied. The circle (diamond) indicate the sets that do (do not) fulfil the three restrictive astrophysical constraints considered. Each diamond is composed of three triangles representing the minimum mass $1.178\,M_\odot$ limit (left triangle),  the maximum mass $2.01\,M_\odot$ (right triangle) limit, and the GW170817 restriction (bottom triangle). The colour green (red) indicates if the corresponding constraint is (is not) satisfied. Although related to specific astronomical observations, these filters can be revealing since they are qualitative indicators of the DM effects on the maximum and minimum mass and on the radius of admixed NSs. The circles in Figs.~\ref{fig:mapa_soft} and \ref{fig:mapa_stiff} indicate parameter sets fulfilling the three constraints. The colour bar indicates the value of the corresponding $M_\mathrm{eq}$ for each EoS set. To simplify the classification and visualisation of the results, we present with dark violet, assigned to $M_\mathrm{eq}=0$, the cases in which no halo branch appears in the mass-radius curve of stable solutions, meaning that $R_\mathrm{eq}$ does not exist. Panels (a), $(b)$ and $(c)$ of Figs.~\ref{fig:mapa_soft} and \ref{fig:mapa_stiff} show the upper left corner excluded according to astrophysical constraints on NSs. This zone corresponds to weakly SIDM or low $y=m_\chi/m_\phi$ values.  Depending on the hadronic EoS and the fraction $f_\chi$ chosen, the excluded region corresponds to $0.2~<~y~<~0.6$  (\textit{stiff} EoS and $f_\chi=0.02$) or  $0.2~<~y~<~10$ (\textit{soft} EoS and $f_\chi=0.1$). In the colour map area, it can be observed that most of the regions (dark purple zone) correspond to EoS with predominantly $\chi$-core branches, as $M_\mathrm{eq}$ becomes remarkably small or absent. Some sets in the intermediate $m_\phi$-low $m_\chi$ region (corresponding to the colours greenish and yellow in the colour map) produce EoS with a non-negligible $\chi$-halo branch contribution. As we already noted in the mass-radius relationships (Figs.~\ref{fig:mraio_soft} and \ref{fig:mraio_stiff}), in the most extreme cases, the $M_\mathrm{eq}$ reach high enough values that one or both objects involved in the merger that produced the GW170817 event could have been $\chi$-halo objects. Although the sampled values of $f_\chi$ differ by approximately a factor $5$, it can be seen in the \textit{soft} case, that a larger value for this parameter prevents compliance of the more restrictive constraint of $2.01\,M_\odot$, producing less massive stellar configurations. On the other hand, due to the higher masses generally achieved by stiff EoSs, our \textit{stiff} models fulfil the maximum mass constraint for almost all studied sets and therefore exclude fewer sets than the \textit{soft} models in this high \mbox{$m_\phi$-low $m_\chi$} region.

\begin{figure*}
\centering
\includegraphics[width=1\linewidth]{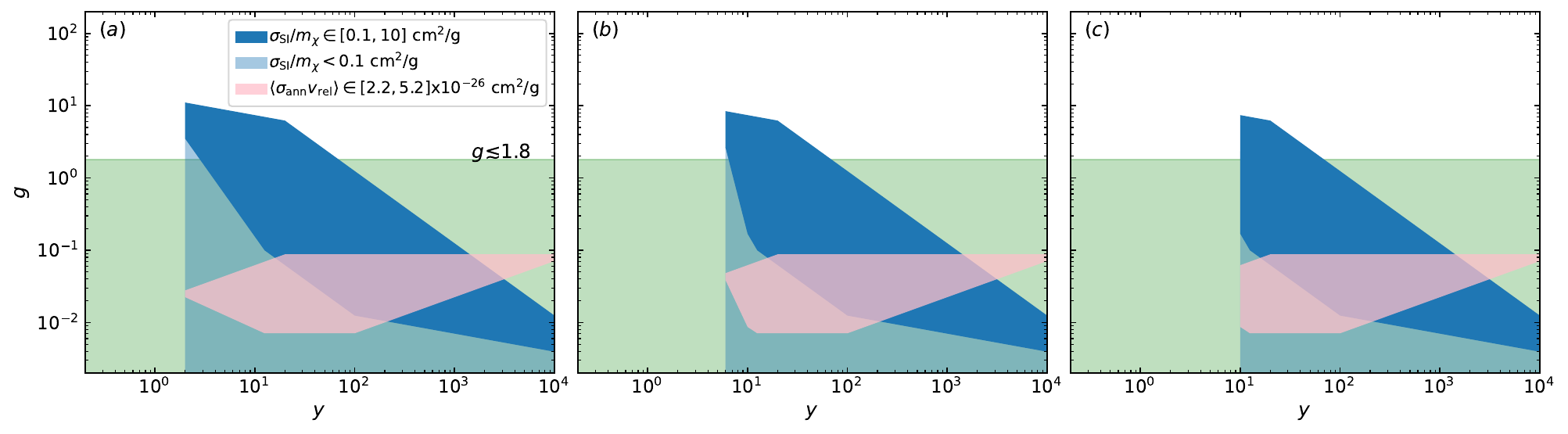}
\caption{Constraints on the generic coupling constant $g$ with respect to $y\equiv m_\chi/m_\phi$. We present the EoSs satisfying the astrophysical constraints (circles in Figs.~\ref{fig:mapa_soft}) for \textit{soft} hadronic EoSs, and different values of $f_\chi=0.02, 0.05, 0.1$ in panels $(a)$, $(b)$ and $(c)$, respectively. The dark and light blue areas are generated by considering Eq.~\eqref{eq:sigma_si} in the interval  $\sigma_\mathrm{SI}/m_\chi \in [0.1, 10]$~cm$^2$/g and the most restrictive limit, $\sigma_\mathrm{SI}/m_\chi <0.1$~cm$^2$/g, respectively. The light pink region represents the assumed DM freeze-out in the range \mbox{$2.2 \times 10^{-26} \lesssim \langle \sigma_\mathrm{ann} v_\mathrm{rel}\rangle \lesssim 5.2 \times 10^{-26}$~cm$^3$/s} \protect\citep{steigman_PhysRevD.86.023506}. The green region indicates the $g<1.8$ constraint arising from Eq.~\eqref{eq:Gamma_decay}.}
\label{fig:sigmaxx_soft}
\end{figure*}

\begin{figure*}
\centering
\includegraphics[width=1\linewidth]{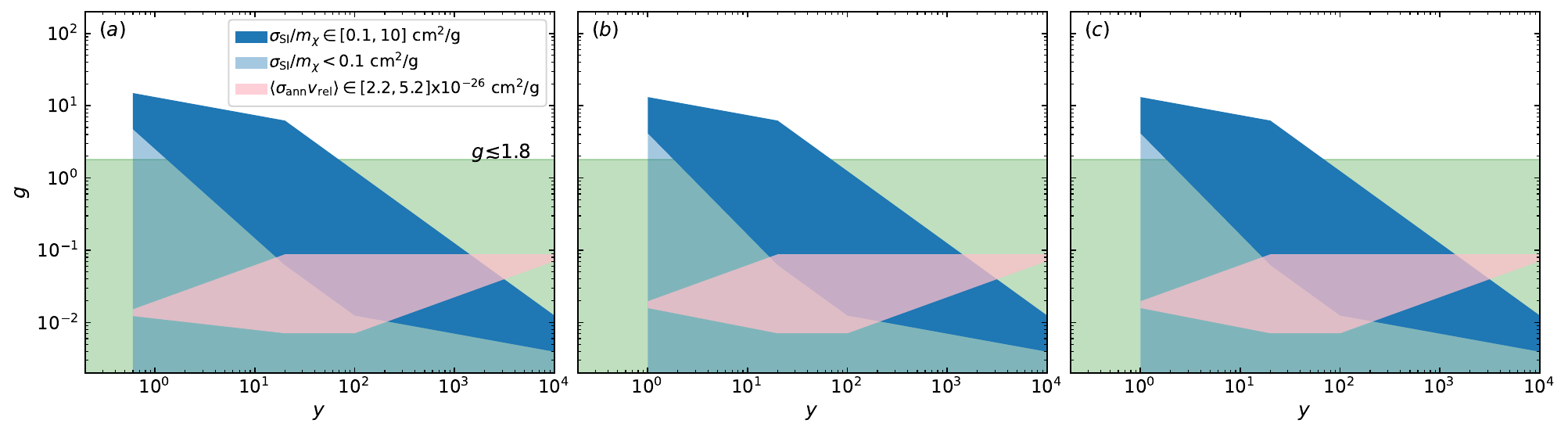}
\caption{Same as Fig.~\ref{fig:sigmaxx_soft} but for \textit{stiff} hadronic EoSs.}
 \label{fig:sigmaxx_stiff}
\end{figure*}

Finally, within the framework of the DM paradigm, in Fig.~\ref{fig:sigmaxx_soft} (soft hadronic EoS) and Fig.~\ref{fig:sigmaxx_stiff} (stiff hadronic EoS) we study the SI generic coupling constant, $g$ - interaction parameter, $y$, plane, considering the constraints on $\sigma_\mathrm{SI}/m_\chi$ and $\langle \sigma_\mathrm{ann} v_\mathrm{rel}\rangle$, and the upper bound of g, given by Eq.~\eqref{eq:Gamma_decay}. For $\sigma_\mathrm{SI}/m_\chi$, we apply \mbox{$\sigma_\mathrm{SI}/m_\chi \lesssim 0.1, 1.0, 10$~cm$^2$/g} corresponding to order-of-magnitude values coming from Bullet-type massive clusters \citep{Robertson:2016wdt}, the Abell cluster galaxies \citep{Kahlhoefer:2015oti} and dwarf galaxies \citep{Hayashi:2021pdm}, respectively; the thermally averaged self-annihilation cross-section from the cosmological DM freeze-out is taken to be \mbox{$2.2 \times 10^{-26}$~cm$^3$/s $\lesssim \langle \sigma_\mathrm{ann} v_\mathrm{rel}\rangle \lesssim 5.2 \times 10^{-26}$~cm$^3$/s} \citep{steigman_PhysRevD.86.023506}. In both figures, panels $(a)$, $(b)$ and $(c)$ correspond to \mbox{$f_{\chi} = 0.02, 0.05, 0.1$}, respectively. The dark blue contoured region in the $g$-$y$ plane is constructed considering the $\sigma_\mathrm{SI}/m_\chi$ in the prescribed uncertainty range $[0.1,10]$~cm$^2$/g, and using the constraints shown in Fig.~\ref{fig:mapa_soft} and \ref{fig:mapa_stiff}, for the mediator mass, the dark fermion mass, considering the restrictions coming from the multi-messenger astronomy of NSs. We colour with light blue the region where $\sigma_\mathrm{SI}/m_\chi <0.1$~cm$^3$/s in accordance with the Bullet cluster constraint. In addition, we present in light pink the constraint for $\langle \sigma_\mathrm{ann} v_\mathrm{rel}\rangle$, also considering the restrictions in $m_\phi$ and $m_\chi$ coming from NS observations. In green colour we present the region $g \lesssim 1.8$, arising from Eq.~\eqref{eq:Gamma_decay}.

In order to establish these constraints over $g$ and $y$ magnitudes, we determine the intersection between the most restrictive limit of $\sigma_{SI}/m_{\chi}<0.1$ and the pink area given by the $\langle \sigma_\mathrm{ann} v_\mathrm{rel}\rangle$ restriction. Although the coloured areas are slightly larger in Fig.~\ref{fig:sigmaxx_stiff} than in Fig.~\ref{fig:sigmaxx_soft}, it is possible to restrict the strength parameter $y$ in the range \mbox{$2 \lesssim y \lesssim 200$}, while the generic coupling constant is allowed to vary between  $0.01\lesssim g \lesssim 0.1$. When considering results for the stiff hadronic EoS, Fig.~\ref{fig:sigmaxx_stiff}, allowed $y$ values are shifted to the lower side with the lowest $y\sim 0.5$ for the $f_\chi=0.02$ case. 

\section{Conclusions} \label{sec:conclusions}

We have studied the impact of the presence of fermionic SIDM on several macroscopic properties of NSs such as gravitational mass, radius and dimensionless tidal deformability. Our study is performed using the two-fluid formalism in which baryonic and dark components interact only gravitationally. To describe the baryonic matter, we have used two extreme cases of GPP EoS i.e. a \textit{soft} and a \textit{stiff} EoS compatible with both low-density chiral EFT calculations and modern NS astronomical observations. It is important to emphasise that the results presented here are intended to be representative of a wide variety of EoS that fall between these two extreme cases. A different choice of \textit{soft} and \textit{stiff} EoS would not qualitatively change the conclusions drawn from our work. The DM EoS was modelled considering a Fermi gas of self-interacting massive particles. Different families of admixed NSs considering a fixed fraction of DM were built through the integration of the TOV equations within the two-fluid formalism.

We have built the mass-radius and tidal~deformability-mass relationships for the different families of stellar configurations with a dark component obtained. The DM parameters associated with the SIDM model considered were varied in a wide range of values to obtain the most general and representative results possible for our model. From the different sets of admixed EoS, we have set a filter to select only those admixed NSs satisfying the current NS constraints. This filtering process allowed us not only to study the different possible NS configurations but also to establish updated constraints over the SIDM parameters.

The results on the mass-radius plane suggest that the effect of the SIDM on admixed NSs could be quite relevant, producing long branches of halo-type objects and/or considerably decreasing the mass and radius of these compact objects. However, when the current astrophysical restrictions of NSs are applied as filters, much of the studied EoSs should be discarded, so these constraints could effectively help improve the DM models. In the cases where the available constraints from NS observations are fulfilled, the effect of fermionic SIDM is more evident for compact low-mass objects producing branches of configurations with approximately constant gravitational mass and a variable radius which is larger than the purely baryonic NS would have, the so-called $\chi$-halo branch. In particular, our results do not discard the possibility that the GW170817 event could be produced by at least one \mbox{$\chi$-halo} object. 

Interestingly, in some cases, we found that the inclusion of a DM component displaces the radius of the obtained configurations towards lower values and contributes, in the case of \textit{stiff} EoS, to satisfy the constraint of the GW170817 event. On the contrary, in the case of the \textit{soft} hadronic EoS, this change in radius could cause the curves to fall outside of the GW190425 restriction. This behaviour is also evident in the tidal deformability-mass plane, where the DM contribution leads to a decrease in $\Lambda$ and, thus, helps satisfy the GW170817 constraint in this plane in the \textit{stiff} scenario.

In addition, we have explored the impact of the presence of SIDM on the NS structure, looking in particular at the change in the Love number $k_2$ which can be estimated indirectly from GW data emitted during binary NS mergers. We find that low compactness admixed NSs present the best scenario for testing the presence of fermionic SIDM in NSs.

The characterisation and classification of the $\chi$-halo and $\chi$-core branches based on the astrophysical constraints allow us to set the most restrictive constraints when studying the DM effects on NSs. In addition to the thoroughly studied maximum mass constraint, the impact of DM on the branches of $\chi$-halo and on the value of $M_\mathrm{eq}$ could potentially be in tension with the minimum mass constraint and the mass and radius constraints coming from GW170817. In this regard, future studies, in the context of potential future observations, should not lose sight of this possible behaviour of admixed NSs.

Considering all the admixed EoSs constructed along with the current astrophysical constraints, we have studied and constrained the parameter space of our fermionic SIDM model. We found that weakly interacting fermionic SIDM -corresponding to low values for $m_\chi$ and high values for $m_\phi$ -i.e., low values for $y$- is excluded for admixed NSs. This effect occurs due to a excessively massive $M_\mathrm{eq}$ -thereby failing to satisfy \mbox{$M_\mathrm{min} < 1.178 \,M_\odot$} or GW170817 - or having a maximum mass $M_\mathrm{max}$ that is too small, less than $2.01\,M_\odot$. In more detail, we have seen that the SIDM EoS constraints cover a wider range in the $m_\phi-m_\chi$ plane if a \textit{soft} hadronic EoS is considered. In this scenario, $m_\phi \gtrsim 10$ MeV is excluded for \mbox{$m_\chi \lesssim 10^3$ MeV}. However, when considering the \textit{stiff} hadronic EoS, this situation undergoes a quantitative change. In this scenario, the constraints coming from the observations of NSs are less stringent, again ruling out DM candidates with masses $m_\chi$ of a few hundred MeV if $m_\phi \gtrsim 10$ MeV. Notably, in this case, the results are almost independent of the DM fraction, $f_\chi$.

The constraints on the values of DM SI cross-sections, obtained from the dynamics of Bullet and Abell clusters, dwarf galaxies, and cosmological DM freeze-out intervals, have been combined with multi-messenger NS astronomy to constrain fermionic SIDM. This wealth of information helps us constrain the $g$-$y$ parameter space effectively. Specifically, we find that, when considering the most restrictive upper limit condition, $\sigma_\mathrm{SI}/m_\chi <0.1$~cm$^3$/s, along with the range of DM freeze values for $\langle \sigma_\mathrm{ann} v_\mathrm{rel}\rangle$, the allowed $g$-$y$ region falls within  $0.01 \lesssim g \lesssim0.1$ and $0.5 \lesssim y \lesssim 200$.

In a recent study by \citet{shirke2023rmodes}, similar SIDM constraints were derived, focusing solely on the DM vector interaction strength and based on only one of the conditions we have imposed, namely, the maximum mass values of NSs. Additionally, they worked within the traditional single-fluid TOV formalism. The updated constraints we present here, considering NS observations and SIDM cross-section restrictions simultaneously, can provide complementary and tighter constraints.

Finally, our results indicate that measuring the mass-radius of a low mass NS potentially in the $\chi$-halo branch or detecting a GW NS merger event with determinations of $\Lambda$ and $k_2$ could offer insights into the presence of DM in NSs. As already mentioned, there are other observable quantities that could also indicate DM presence in compacts objects, such as gravitational waves from NS oscillations \citep{shirke2023rmodes}, explosive kilonovae events \citep{Bramante:2014bos} or X-ray pulse profiles \citep{Miao:2022dma}. We anticipate that this diverse range of observable predictions, coupled with the wealth of data expected from future experimental and observational projects, will contribute significantly to advancing our understanding of both the fundamental nature of DM and the internal structure and composition of NSs.

\section*{Acknowledgements} 
We would like to thank the anonymous referee for her/his comments that have contributed to improve the quality of this work. This work has been supported by Junta de Castilla y Le\'on SA096P20 and Spanish MICIN grant PID2019-107778GB-I00 and PID2022-137887NB-I00. M.M, M.G.O and I.F.R-S thank CONICET and UNLP for financial support under grants PIP-0169 and 11/G187. M.G.O and I.F.R-S are partially supported by the National Science Foundation (USA) under Grant PHY-2012152. I.F.R-S is partially supported by PICT grant 2019-0366 from ANPCyT and PIBAA grant 0724 from CONICET (Argentina).

\section*{Data Availability}
The computed data presented and discussed in this paper will be shared upon reasonable request.

\bibliographystyle{mnras}
\bibliography{dark} 








\bsp	
\label{lastpage}
\end{document}